\documentclass[twocolumn,showpacs,preprintnumbers, 
               superscriptaddress]{revtex4}
\usepackage{graphicx, fancybox}
\usepackage{amsmath,amssymb}

\begin{document}


\title{
Relation between baryon number fluctuations and 
experimentally observed proton number fluctuations 
in relativistic heavy ion collisions 
}

\author{Masakiyo Kitazawa}
\email{kitazawa@phys.sci.osaka-u.ac.jp}
\affiliation{
Department of Physics, Osaka University, Toyonaka, Osaka 560-0043, Japan}

\author{Masayuki Asakawa}
\email{yuki@phys.sci.osaka-u.ac.jp}
\affiliation{
Department of Physics, Osaka University, Toyonaka, Osaka 560-0043, Japan}

\begin{abstract}

We explore the relation between proton and nucleon number
fluctuations in the final state in 
relativistic heavy ion collisions.
It is shown that the correlations between the isospins of 
nucleons in the final state are almost negligible 
over a wide range of collision energy.
This leads to a factorization of the distribution function 
of the proton, neutron, and their antiparticles
in the final state with binomial distribution functions.
Using the factorization, we derive formulas to determine 
nucleon number cumulants, which are not direct 
experimental observables, from proton number fluctuations,
which are experimentally observable in event-by-event analyses.
With a simple treatment for strange baryons,
the nucleon number cumulants 
are further promoted to the baryon number ones.
Experimental determination of the baryon number cumulants 
makes it possible to compare various theoretical studies on 
them directly with experiments.
Effects of nonzero isospin density on this formula are
addressed quantitatively. It is shown that the effects
are well suppressed over a wide energy range.

\end{abstract}

\date{\today}

\maketitle

\section{Introduction}

Now that the observation of the quark-gluon matter in 
relativistic heavy ion collisions is established for 
small baryon chemical potential ($\mu_{\rm B}$) \cite{RHIC},
a challenging experimental subject following this 
achievement is to reveal the global structure of the QCD phase 
diagram on the temperature ($T$) and $\mu_{\rm B}$ plane.
In particular, finding the QCD critical point(s), 
whose existence is predicted by various theoretical
studies \cite{Asakawa:1989bq,Stephanov:2004wx}, is one 
of the most intriguing problems.
Since the $\mu_{\rm B}$ of the hot medium created by heavy-ion
collisions can be controlled by varying the collision 
energy per nucleon pair, $\sqrt{s_{\rm NN}}$, the $\mu_{\rm B}$ dependence of the 
nature of QCD phase transition should be observed as the 
$\sqrt{s_{\rm NN}}$ dependence of observables.
An experimental project to explore such signals 
in the energy range $10{\rm GeV}\lesssim \sqrt{s_{\rm NN}} 
\lesssim 200{\rm GeV}$, which is called the energy 
scan program, is now ongoing at the Relativistic 
Heavy Ion Collider (RHIC) \cite{Aggarwal:2010wy,Mohanty:2011nm}.
Experimental data which will be obtained in future
experimental facilities designed for lower beam-energy 
collisions will also provide important 
information on this subject \cite{Bleicher:2011jk}.

Observables which are suitable to analyze 
bulk properties of the matter around the phase boundary 
of QCD in heavy ion collisions are fluctuations 
\cite{Koch:2008ia}.
Experimentally, fluctuations are measured through 
event-by-event analyses \cite{Aggarwal:2010wy}.
Theoretically, it is predicted that some of them, 
including higher-order cumulants, are sensitive to 
critical behavior near the QCD critical point 
\cite{Stephanov:1998dy,Hatta:2003wn,Stephanov:2008qz,
Athanasiou:2010kw,Fraga:2011hi}, 
and/or locations on the phase diagram, especially 
on which side the system is, 
the hadronic side or the quark-gluon side
\cite{Asakawa:2000wh,Jeon:2000wg,Koch:2005vg,Ejiri:2005wq,
Asakawa:2009aj,Friman:2011pf,Stephanov:2011pb}.

Among the fluctuation observables, those of conserved charges 
are believed to possess desirable properties 
to probe the phase structure in relativistic heavy ion 
collisions.
One of the advantages of using the conserved charges 
is that the characteristic times for the variation of their local 
densities are longer than those for non-conserved ones, because the
variation of the local densities of conserved charges are
achieved only through 
diffusion \cite{Asakawa:2000wh,Jeon:2000wg}.
The fluctuations of the former
thus can better reflect fluctuations generated 
in earlier stages of fireballs, when the rapidity coverage
is taken sufficiently large.
From a theoretical point of view, an important property 
of the conserved charges is that one can define the 
operator of a conserved charge, $Q$, as a Noether current. 
Moreover, their higher-order cumulants, $\langle \delta Q^n 
\rangle_c$, are directly related to the grand canonical 
partition function $Z(\mu) = {\rm Tr} e^{-\beta (H-\mu Q)}$ 
as 
\begin{align}
\langle \delta Q^n \rangle_c 
= T^n \frac{ \partial^n \log Z(\mu) }{ \partial \mu^n },
\label{eq:Q}
\end{align}
with $H$ and $\mu$ being the hamiltonian and the chemical 
potential associated with $Q$, respectively.
These properties make the analysis of cumulants of 
conserved charges well defined and feasible
in a given theoretical framework.
For example, they can be measured in lattice QCD Monte 
Carlo simulations \cite{Gavai:2010zn,
Schmidt:2010xm,Mukherjee:2011td,Borsanyi:2011sw,Bazavov:2012jq}. 
The relation Eq.~(\ref{eq:Q}) also provides an intuitive 
interpretation for the behavior of higher-order cumulants
of conserved charges.
For instance, the third-order cumulant of the net baryon number, 
$N_{\rm B}^{\rm (net)}$, satisfies 
$\langle (\delta N_{\rm B}^{\rm (net)})^3 \rangle_c = T
\partial \langle (\delta N_{\rm B}^{\rm (net)})^2 \rangle_c /
\partial \mu_{\rm B} $.
This formula means that 
$\langle (\delta N_{\rm B}^{\rm (net)})^3 \rangle_c$
changes its sign around the phase boundary on the 
$T$-$\mu_{\rm B}$ plane where the baryon number 
susceptibility $\langle (\delta N_{\rm B}^{\rm (net)})^2 \rangle$ 
has a peak structure \cite{Asakawa:2009aj}.
The change of the sign of observables like this will serve as 
a clear experimental signal 
\cite{Asakawa:2009aj,Friman:2011pf,Stephanov:2011pb}.

QCD has several conserved charges, such as baryon and 
electric charge numbers and energy.
Among these conserved charges, theoretical studies 
suggest that the cumulants of the baryon number have 
the most sensitive dependences on the phase transitions 
and phases of QCD.
In order to see this feature,
let us compare the baryon number
cumulants with the electric charge ones.
First, the baryon number fluctuations show the 
critical fluctuations associated with the QCD critical 
point more clearly.
Although the baryon and electric charge number fluctuations 
diverge with the same critical exponent around the critical 
point, it should be remembered that this does not mean similar clarity of 
signals for the critical enhancement in experimental studies.
Fluctuations near the critical point are generally separated
into singular and regular parts, and only the former 
diverges with the critical exponent.
The singular part of the electric charge fluctuations
is relatively suppressed compared to the baryon number ones,
because the formers contain the isospin number fluctuations
which are regular near the critical point \cite{Hatta:2003wn}.
The additional regular contribution makes the experimental
confirmation of the enhancement of the singular part difficult, 
and this tendency is more pronounced in 
higher-order cumulants 
\cite{Asakawa:2009aj}.
While it is known that the proton number fluctuations in 
the final state also reflect the critical enhancement 
near the critical point \cite{Hatta:2003wn}, as we will show 
later the baryon number fluctuations are superior to this 
observable, too, in the same sense.
Second, the ratios of baryon number cumulants \cite{Ejiri:2005wq} 
behave more sensitively to the difference of phases, i.e.,
hadrons, or quarks and gluons.
This is because the ratios are dependent on the magnitude of 
charges carried by the quasi-particles composing the 
state \cite{Asakawa:2000wh,Jeon:2000wg,Ejiri:2005wq}, 
while the charge difference between hadrons 
and quarks is more prominent 
in the baryon number.

Experimentally, however, the baryon number fluctuations 
are not directly observable, because chargeless 
baryons, such as neutrons, cannot be detected by
most detectors.
Proton number fluctuations can be measured 
\cite{Aggarwal:2010wy,Mohanty:2011nm}, 
and recently its cumulants have been compared with
theoretical predictions for baryon number cumulants.
Indeed, in the free hadron gas in equilibrium
the baryon number cumulants are approximately twice the 
proton number ones, because 
the baryon number cumulants in free gas are simply given by
the sum of those for all baryons, and the baryon 
number is dominated by proton and neutron numbers in 
the hadronic medium relevant to relativistic heavy ion collisions.
In general, however, these cumulants behave differently.
In fact, we will see later that the non-thermal effects
which exists in baryon number cumulants are strongly suppressed in 
the proton number ones.

In heavy ion collisions, because of the dynamical evolution
the medium at kinetic freezeout is not completely in the 
thermal equilibrium.
The original ideas to exploit fluctuation observables
as probes of primordial properties of fireballs
\cite{Asakawa:2000wh,Jeon:2000wg} are concerned 
with this non-thermal effect encoded in the final 
state as a hysteresis of the time evolution.
To observe such effects, it is highly desirable to measure 
baryon number cumulants that is expected to retain more
effects of the phase transition and the singularity around
the critical point.
The experimental determination of baryon number cumulants 
also makes the comparison between experimental and 
theoretical studies more robust, since many theoretical works 
including lattice QCD simulations are 
concerned with the baryon number cumulants, not the proton 
number ones.

In Ref.~\cite{Kitazawa:2011wh}, the authors of the present 
paper have argued that, whereas the baryon number 
cumulants are not the direct experimental observables
as discussed above, 
they can be determined in experiments by only using 
the experimentally measured proton number fluctuations 
for $\sqrt{s_{\rm NN}}\gtrsim10{\rm GeV}$.
The key idea is that isospins of nucleons 
in the final state are almost completely randomized and 
uncorrelated, because of reactions of nucleons with 
thermal pions in the hadronic stage, as will be elucidated 
in Sec.~\ref{sec:B}. 
This leads to the conclusion 
that, when $N_{\rm N}$ nucleons exist in a phase 
space of the final state, the probability that $N_p$ nucleons 
among them are protons follows the binomial distribution.
More generally, the probability distribution that $N_p$ 
protons, $N_n$ neutrons, $N_{\bar p}$ anti-protons, 
and $N_{\bar n}$ anti-neutrons are found in the final state 
in a phase space is factorized as
\begin{align}
\lefteqn{
{\cal P}_{\rm N}( N_p , N_n , N_{\bar p} , N_{\bar n} )
}
\nonumber \\
&= 
{\cal F}(N_{\rm N},N_{\bar{\rm N}})
B_r(N_p;N_{\rm N}) B_{\bar r}(N_{\bar p};N_{\bar{\rm N}}),
\label{eq:P}
\end{align}
where the nucleon and anti-nucleon numbers are 
$N_{\rm N}=N_p+N_n$ and $N_{\bar{\rm N}}=N_{\bar p}+N_{\bar n}$, 
respectively, and 
\begin{align}
B_r(k;n) =  \frac{ n! }{ k! (n-k)! } r^k (1-r)^{n-k}
\label{eq:B}
\end{align}
is the binomial distribution function with probabilities
$r=\langle N_p\rangle/\langle N_{\rm N}\rangle$ and 
$\bar{r}=\langle N_{\bar p}\rangle/\langle N_{\bar{\rm N}}\rangle$.
The function ${\cal F}(N_{\rm N},N_{\bar{\rm N}})$ describes 
the distribution of nucleons and anti-nucleons and the correlation
between them in the final state, 
which are determined by the dynamical history of fireballs.
Using the factorization Eq.~(\ref{eq:P}), one can obtain
formulas to represent the (anti-)nucleon number cumulants by 
the (anti-)proton number ones, and vice versa;
whereas the neutron number is not determined by experiments, 
this missing information can be reconstructed with the knowledge
for the distribution function, Eq.~(\ref{eq:P}).
The (anti-)nucleon number in Eq.~(\ref{eq:P}) can 
further be promoted to the (anti-)baryon number 
in practical analyses with a 
simple treatment for strange baryons to a good approximation.
These formulas enable to determine the baryon number 
cumulants solely with the experimentally measured proton 
number fluctuations, and, as a result, to obtain insights into 
the present experimental results on the proton number 
cumulants.

The main purpose of the present paper is to elaborate the 
discussion in Ref.~\cite{Kitazawa:2011wh} with some extensions. 
In Ref.~\cite{Kitazawa:2011wh} the formulas are 
derived only for isospin symmetric medium.
In the present study we extend them to incorporate 
cases with nonzero isospin densities.
With the extended relations, it is shown that the effect 
of nonzero isospin density is well suppressed for 
$\sqrt{s_{\rm NN}}\gtrsim10{\rm GeV}$.
The procedures of the manipulations and discussions
omitted in Ref.~\cite{Kitazawa:2011wh} are also addressed 
in detail.

In the next Section, we show that the factorization 
Eq.~(\ref{eq:P}) is well applied to the nucleon and baryon 
distribution functions in the final state in heavy ion collisions.
We then derive formulas to relate baryon and proton number 
cumulants in Sec.~\ref{sec:cumulant}.
In Sec.~\ref{sec:discussion}, we discuss the recent 
experimental results at STAR \cite{Aggarwal:2010wy,Mohanty:2011nm} 
using the results 
in Sec.~\ref{sec:cumulant}, and possible extensions of 
our results.
The final section is devoted to a short summary.

Throughout this paper, we use $N_X$ to represent 
the number of particles $X$ leaving the system after each 
collision event, where $X=p$, $n$, N, and B represent 
proton, neutron, nucleon, and baryon, respectively, and 
their anti-particles, $\bar p$, $\bar n$, $\bar{\rm N}$, 
and $\bar{\rm B}$.
The net and total numbers are defined as 
$N_X^{\rm (net)} = N_X - N_{\bar{X}}$ and 
$N_X^{\rm (tot)} = N_X + N_{\bar{X}}$, respectively.

\section{Distribution function for proton and neutron numbers}
\label{sec:B}

In this section, we discuss the time evolution of the proton 
and neutron number distributions in the hadronic medium 
generated by relativistic heavy ion collisions, and show that 
the nucleon distribution in the final state in a phase 
space is factorized as in Eq.~(\ref{eq:P}) at sufficiently
large $\sqrt{s_{\rm NN}}$.
In Sec.~\ref{sec:B:free}, as a preliminary example we show 
that Eq.~(\ref{eq:P}) is applicable to the 
equilibrated free hadron gas in the ranges of $T$ 
and $\mu_{\rm B}$ relevant to relativistic heavy ion collisions.
We then extend the argument to the distribution function 
in the final state in relativistic heavy ion collisions 
in Sec.~\ref{sec:B:HIC}.

\subsection{Free hadron gas in equilibrium}
\label{sec:B:free}

Let us first consider nucleons in the free hadron gas in 
equilibrium.
For $T$ and $\mu_{\rm B}$ which are relevant to 
relativistic heavy ion collisions, the nucleon mass $m_{\rm N}$ 
satisfies $m_{\rm N} - |\mu_{\rm B}| \gg T$.
One thus can apply the Boltzmann approximation
for the distribution functions of nucleons.
The number of particles in a phase space, $N$, which obey
Boltzmann statistics is given by the Poisson distribution,
\begin{align}
P_{\lambda}(N) = \frac{ e^{-\lambda} \lambda^N }{ N! },
\end{align}
with the average $\lambda = \langle N \rangle 
= \sum_N N P_{\lambda}(N)$.
Accordingly, the probability to find $N_p$ ($N_{\bar p}$) 
protons (anti-protons) and $N_n$ ($N_{\bar n}$) neutrons 
(anti-neutrons) in the phase space is given by the product of
the Poisson distribution functions,
\begin{align}
\lefteqn{{\cal P}_{\rm HG}( N_p , N_n , N_{\bar p} , N_{\bar n} )}
\nonumber \\
&= 
P_{\langle N_p \rangle} (N_p)
P_{\langle N_n \rangle} (N_n)
P_{\langle N_{\bar p} \rangle} (N_{\bar p})
P_{\langle N_{\bar n} \rangle} (N_{\bar n}).
\label{eq:P_HG1}
\end{align}

The product of two Poisson distribution functions 
satisfies the identity,
\begin{align}
\lefteqn{ P_{\lambda_1}(N_1) P_{\lambda_2}(N_2) }
\nonumber \\
&= P_{\lambda_1+\lambda_2}(N_1+N_2) 
B_{\lambda_1/(\lambda_1+\lambda_2)}( N_1 ; N_1+N_2 ),
\label{eq:PP=PB}
\end{align}
where $B_r(k;n)$ is the binomial distribution function
Eq.~(\ref{eq:B}).
Using Eq.~(\ref{eq:PP=PB}), 
Eq.~(\ref{eq:P_HG1}) is rewritten as 
\begin{align}
\lefteqn{ {\cal P}_{\rm HG}( N_p , N_n , N_{\bar p} , N_{\bar n} ) }
\nonumber \\
&= 
P_{\langle N_{\rm N} \rangle} (N_{\rm N})
P_{\langle N_{\bar{\rm N}} \rangle} (N_{\bar{\rm N}})
B_r(N_p;N_{\rm N}) B_{\bar r}(N_{\bar p};N_{\bar{\rm N}}),
\label{eq:P_HG}
\end{align}
where $N_{\rm N}=N_p+N_n$ and $N_{ \bar{\rm N}}=N_{\bar p}+N_{\bar n}$ 
are the nucleon and anti-nucleon numbers, respectively, and 
$r= \langle N_p \rangle / \langle N_{\rm N} \rangle$ and 
$\bar{r}= \langle N_{\bar p} \rangle / \langle N_{\bar{\rm N}} \rangle$.
Equation~(\ref{eq:P_HG}) shows that the distribution of 
nucleons in the free hadron gas is factorized using 
binomial functions as in Eq.~(\ref{eq:P}) with
\begin{align}
{\cal F} ( N_{\rm N},N_{\bar{\rm N}} )
= P_{\langle N_{\rm N} \rangle} (N_{\rm N})
P_{\langle N_{\bar{\rm N}} \rangle} (N_{\bar{\rm N}}).
\label{eq:F_HG}
\end{align}

The appearance of the binomial distribution functions 
in Eq.~(\ref{eq:P_HG}) is understood as follows. 
When one finds a nucleon in the hadron gas, the probability 
that the nucleon is a proton is $r$.
The isospins of all nucleons found in the phase space, moreover, 
are not correlated with one another as a consequence of 
Boltzmann statistics and the absence of interactions.
Once $N_{\rm N}$ nucleons are found in the phase space, 
therefore, the probability that $N_p$ particles
are protons is a superposition of independent events
with probability $r$, i.e.,
the binomial distribution.

We note that the above discussion is not applicable 
when the condition $m_{\rm N} - |\mu_{\rm B}| \gg T$, 
required for Boltzmann statistics, is not satisfied.
When quantum correlations of nucleons arising from 
Fermi statistics are not negligible, the isospin of 
each nucleon can no longer be independent.
As long as we are concerned with the range of $T$ and 
$\mu_{\rm B}$ which can be realized by relativistic
heavy ion collisions, however, the condition for the
Boltzmann approximation is 
well satisfied except in very low energy collisions
\cite{Cleymans:1998fq}.

\subsection{Final state in heavy ion collisions}
\label{sec:B:HIC}

Next, we consider the nucleon distribution functions
in the final state in heavy ion collisions.
We show that the nucleon distribution in this case 
is also factorized as in Eq.~(\ref{eq:P}), by demonstrating 
that the isospins of all nucleons in the final state
are random and uncorrelated.

\subsubsection{$\Delta(1232)$ resonance}

The key ingredient to obtain the factorization 
Eq.~(\ref{eq:P}) in the final state in relativistic
heavy ion collisions is ${\rm N}\pi$ reactions
in the hadronic stage mediated by 
$\Delta(1232)$ resonances having the isospin $I=3/2$.
As we will see later, this is the most dominant reaction 
of nucleons in the hadronic medium.
This is because 
i) the cross section of ${\rm N}\pi\to\Delta$ reactions 
exceeds $200{\rm mb}=20{\rm fm}^2$ and is comparable with 
${\rm NN}$ and ${\rm N \bar{N}}$ reactions 
for $P_{\rm lab}\simeq 300{\rm MeV}$ \cite{PDG}, and 
ii) the pion density dominates over those of all other 
particles in the ranges of $T$ and $\mu_{\rm B}$ 
accessible with heavy ion collisions at 
$\sqrt{s_{\rm NN}}\gtrsim 10{\rm GeV}$;
at the top RHIC energy, the density of pions
is more than one order larger than that of nucleons.
We shall show below that these reactions 
frequently take place even after chemical freezeout in 
the hadronic medium during the time evolution of the 
fireballs.

The ${\rm N}\pi$ reactions through $\Delta$ contain
charge exchange reactions, which alter the isospin of
the nucleon in the reaction.
The reactions of a proton to form $\Delta$ 
are:
\begin{align}
p + \pi^+ &\to \Delta^{++} \to p + \pi^+,
\label{eq:D++} 
\\
p + \pi^0 &\to \Delta^+ \to p(n) + \pi^0(\pi^+),
\label{eq:D+}
\\
p + \pi^- &\to \Delta^0 \to p(n) + \pi^-(\pi^0).
\label{eq:D0}
\end{align}
Among these reactions, Eqs.~(\ref{eq:D+}) and (\ref{eq:D0}) 
are responsible for the change of the nucleon isospin.
The ratio of the cross sections of a proton to form 
$\Delta^{++}$, $\Delta^+$, and $\Delta^0$ is $3:1:2$, 
which is determined by the isospin SU(2) symmetry of 
the strong interaction.
The isospin symmetry also tells us that the branching 
ratios of $\Delta^+$ ($\Delta^0$) decaying into the final 
state having a proton and a neutron are $1:2$ ($2:1$).
Using these ratios, one obtains the ratio of the probabilities
that a proton in 
the hadron gas forms $\Delta^+$ or $\Delta^0$ with
a reaction with a thermal pion, and then 
decays into a proton and a neutron, respectively, $P_{p\to p}$ and 
$P_{p\to n}$, as
\begin{align}
P_{p\to p} : P_{p\to n} = 5:4,
\label{eq:P_pn}
\end{align}
provided that the hadronic medium is isospin symmetric
and that the three isospin states of the pion are equally 
distributed in the medium.
Because of the isospin symmetry of the strong interaction 
one also obtains the same conclusion for neutron reactions:
\begin{align}
P_{n\to n} : P_{n\to p} = 5:4.
\label{eq:P_np}
\end{align}
Similar results are also obtained for anti-nucleons.
Equations~(\ref{eq:P_pn}) and (\ref{eq:P_np}) show that 
these reactions act to randomize the isospin 
of nucleons during the hadronic stage.

\begin{figure}[tbp]
\begin{center}
\includegraphics[width=.49\textwidth]{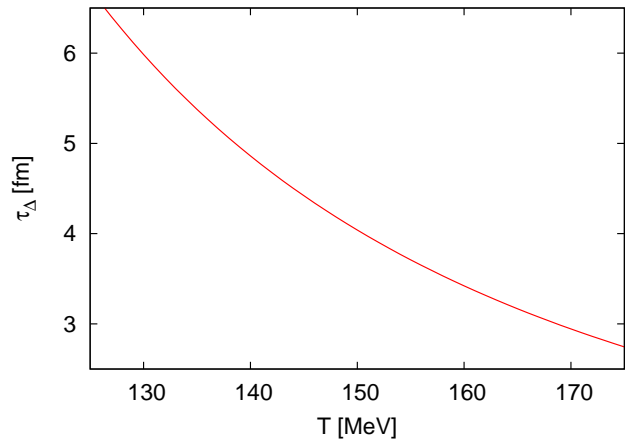}
\caption{
Mean time $\tau_\Delta$ of a rest nucleon to form 
$\Delta^+$ or $\Delta^0$ in the hadronic medium 
as a function of temperature $T$.
}
\label{fig:tau}
\end{center}
\end{figure}

\subsubsection{Mean time}

Next, let us estimate the mean time of these reactions.
Assuming that pions are thermally distributed,
the mean time $\tau_\Delta$ of a nucleon at rest in
the medium to undergo a reaction Eq.~(\ref{eq:D+}) or 
(\ref{eq:D0}) is given by
\begin{align}
\tau_\Delta^{-1} = \int \frac{d^3 k_\pi}{(2\pi)^3} 
\sigma(E_{\rm c.m.}) v_\pi n(E_\pi),
\label{eq:tau}
\end{align}
with the Bose distribution function $n(E)= ( e^{E/T}-1)^{-1}$,
the pion momentum $k_\pi$,
the pion velocity $v_\pi=k_\pi/E_\pi$, $E_\pi=\sqrt{m_\pi^2+k_\pi^2}$, 
and the pion mass $m_\pi$.
$\sigma(E_{\rm c.m.})$ is the sum of the cross sections for N$\pi$ 
reactions producing $\Delta^+$ and $\Delta^0$ for 
the center-of-mass energy 
$E_{\rm c.m.}=[ (m_{\rm N}+E_\pi)^2 - k_\pi^2]^{1/2}$ 
with the nucleon mass $m_{\rm N}$.
For the cross section $\sigma(E_{\rm c.m.})$, we 
assume that the peak corresponding to $\Delta(1232)$ 
resonance is well reproduced by the Breit-Wigner form, 
\begin{align}
\sigma(E_{\rm c.m.}) = \sigma_\Delta
\frac{ \Gamma^2/4 }{ (E_{\rm c.m.}-m_\Delta)^2 + \Gamma^2/4 },
\end{align}
which is a sufficient approximation for our purpose.
Here, we use the value of the parameters determined by the N$\pi$
reactions in the vacuum, 
$m_\Delta=1232{\rm MeV}$, $\Gamma=110{\rm MeV}$, 
and $\sigma_\Delta=20{\rm fm}^2$ \cite{PDG}.
The medium effects on the cross section will be discussed later.
Substituting $m_{\rm N}=940{\rm MeV}$ and $m_\pi=140{\rm MeV}$,
one obtains the $T$ dependence of the mean time $\tau_\Delta$ 
presented in Fig.~\ref{fig:tau}.
The figure shows that the mean time is $\tau_\Delta =
3\sim4$fm for $T=150\sim170{\rm MeV}$. 
One can confirm that the mean time hardly changes even 
for moving nucleons in the range of momentum $p\lesssim 3T$ 
by extending Eq.~(\ref{eq:tau}) to cases with
nonzero nucleon momentum.
The lifetime of $\Delta$ resonances is 
$\tau_\Gamma = 1/\Gamma \simeq1.8{\rm fm}$.

The mean time evaluated above is much shorter than the lifetime 
of the hadronic stage in relativistic heavy ion collisions. 
According to a dynamical model analysis for collisions 
at RHIC, nucleons in the hadron phase 
continue to interact for a couple of tens of fm on average 
at midrapidity \cite{Nonaka:2006yn}.
As a result, at the RHIC energy each nucleon in 
a fireball has chances to undergo the charge exchange 
reactions several times in the hadronic stage.

Two remarks are in order here.
First, the above result on the time scales shows that the 
reactions to produce $\Delta$
proceed even after chemical freezeout.
These reactions do not contradict the success of the 
statistical model, which describes the chemical freezeout 
\cite{BraunMunzinger:2003zd}, because chemical freezeout 
is a concept to describe ratios of particle abundances 
such as $\langle N_{\bar p} \rangle / \langle N_p \rangle$
and the above reactions do not alter the average 
abundances in the final state.
The success of the model, on the other hand, indicates that 
creations and annihilations of (anti-)nucleons hardly occur 
after chemical freezeout.
Second, we note that the dynamical model in 
Ref.~\cite{Nonaka:2006yn} uses an equation of states having 
a first order phase transition in the hydrodynamic 
simulations for the time evolution above the critical 
temperature $T_{\rm c}$.
Recently, dynamical simulations have been carried out with 
more realistic equations of states obtained by lattice 
QCD simulations \cite{Schenke:2010nt}.
The lifetime of hadronic 
stage evaluated in these studies is more relevant to this argument.
We, however, note that the qualitative behavior of the 
time evolution seems not sensitive to the
difference in equations of states \cite{Schenke:2010nt}.

While N$\pi$ reactions frequently take place even below 
the chemical freezeout temperature, $T_{\rm chem}$, 
${\rm N}\bar{\rm N}$ annihilatios and productions almost 
terminate at $T_{\rm chem}$.
This is necessary for the success of the thermal model.
For $E_{\rm c.m.}\simeq T$ the cross section 
of the ${\rm N}\bar{\rm N}$ pair annihilation is largest 
among all NN and N$\bar{\rm N}$ reactions.
If nucleons and anti-nucleons are distributed without
correlation, therefore, all NN and N$\bar{\rm N}$ reactions cease
to take place at $T_{\rm chem}$.
This conclusion is, of course, obtained also by evaluating 
the mean time for each reaction using the cross sections 
\cite{PDG} as in Eq.~(\ref{eq:tau}).
After chemical freezeout, 
the only inelastic reactions nucleons go through are
thus Eqs.~(\ref{eq:D+}) and (\ref{eq:D0}), 
and after each reaction the 
nucleon loses its initial isospin information.
Only after repeating the reactions Eq.~(\ref{eq:P_pn}) 
twice, the ratio becomes $41:40$, which is almost even.
If medium effects on the formations and decays of $\Delta$
are negligible, therefore, irrespective of the 
nucleon distribution at the chemical freezeout,
the isospin of nucleons at the kinetic freezeout 
can be regarded random and uncorrelated.
On the other hand, the nucleon number distribution 
can have a deviation from the Boltzmann distribution 
reflecting the dynamical history of fireballs.

Because of the absence of correlations between isospins of 
nucleons in the final state, once $N_{\rm N}$ ($N_{\bar{\rm N}}$) 
nucleons (anti-nucleons) exist in a phase space in the final 
state, their isospin distribution is simply given by the 
binomial one.
This conclusion leads to the factorization Eq.~(\ref{eq:P})
for proton and neutron number distribution in the final state 
for an arbitrary phase space.
In particular, the final state
proton and anti-proton number distribution is written as 
\begin{align}
{\cal G}(N_p,N_{\bar p})
&= \sum_{N_n,N_{\bar n}} 
{\cal P}_{\rm N} ( N_p , N_n , N_{\bar p} , N_{\bar n} )
\nonumber \\
&= \sum_{N_{\rm N},N_{\bar{\rm N}}} 
{\cal F}(N_{\rm N},N_{\bar{\rm N}})
B_r ( N_p;N_{\rm N} ) B_{\bar r} ( N_{\bar p} ; N_{\bar{\rm N}} ).
\label{eq:gg}
\end{align}
Unlike in the simple example in Sec.~\ref{sec:B:free}, the nucleon 
distribution function ${\cal F}(N_{\rm N},N_{\bar{\rm N}})$
in this case is determined by the time evolution of 
fireballs and is not necessary of a thermal or separable form 
as in Eq.~(\ref{eq:F_HG}); 
no specific form for ${\cal F}(N_{\rm N},N_{\bar{\rm N}})$ is assumed 
here or will be assumed in the analyses in Sec. \ref{sec:cumulant}.
What we have used here is the fact that the time scale for 
the exchange of isospins between nucleons and pions is 
sufficiently short compared to the lifetime of hadronic stage 
after the chemical freezeout.
On the other hand, the time scale for the variation of a conserved 
charge in a phase space depends on the form of the phase space,
and can become arbitrary long by increasing the spatial volume.
When the time scale is long, the information of the physics of 
the early stages is encoded in ${\cal F}(N_{\rm N},N_{\bar{\rm N}})$.

\subsubsection{Medium effects}

Next, let us inspect the possibility of medium effects on 
the formation and decay rates of $\Delta$.
In medium, the decay rate of $\Delta$ acquires the statistical factor, 
\begin{align}
\left( 1-f(E_{\rm N}) \right) \left( 1+n(E_\pi) \right),
\label{eq:f}
\end{align}
where $f(E)=(e^{(E-\mu_{\rm B})/T}+1)^{-1}$ is the Fermi distribution 
function and $E_{\rm N}$ and $E_\pi$ are the energies of the nucleon 
and pion produced by the decay, respectively.
The first term in Eq.~(\ref{eq:f}) represents the 
Pauli blocking effect.
At the RHIC energy, since the Boltzmann approximation is 
well applied to nucleons, the Pauli blocking effect is 
suppressed. The Bose factor $(1+n(E_\pi))$ in Eq.~(\ref{eq:f}), on the 
other hand, has a non-negligible contribution since 
$m_\pi\simeq T_{\rm chem}$.
As long as all $n(E_\pi)$ for the three isospin states of the pion 
are the same, however, this factor does not alter the 
branching ratios Eqs.~(\ref{eq:P_pn}) and (\ref{eq:P_np}),
while the factor enhances the decay of $\Delta$.
A possible origin for the variation of $n(E_\pi)$ is the 
isospin density of nucleon number; since the isospin density
is locally conserved, the isospin density of pions is 
affected by the nucleon isospin.
This effect on $n(E_\pi)$ is, however, well suppressed since 
the density of pions is much larger than that of nucleons below 
$T_{\rm chem}$. 
Another possible source which gives rise to a different pion 
distribution is the event-by-event fluctuation of the isospin 
density in the phase space at the hadronization.
It is, however, expected that the effect is well suppressed, 
again because of the large pion density.
One, therefore, can conclude that the medium effect hardly
changes the branching ratios Eqs.~(\ref{eq:P_pn}) and 
(\ref{eq:P_np}).
The same conclusion also applies to the formation rate of 
$\Delta$, since the medium effect on the probabilities of a 
nucleon to undergo reactions Eqs.~(\ref{eq:D++}) - (\ref{eq:D0})
depends only on $n(E_\pi)$.
After all, all medium effects on the ratios 
Eqs.~(\ref{eq:P_pn}) and (\ref{eq:P_np}) are negligible.

When the system has a nonzero isospin density, 
probabilities Eqs.~(\ref{eq:P_pn}) and (\ref{eq:P_np})
receive modifications because the three isospin states of 
the pion are not equally distributed, although this effect is 
not large as will be shown in Sec.~\ref{sec:cumulant:nid}.
Even in this case, however, the only modification to 
the above conclusion is to replace the probabilities 
$r$ and $\bar{r}$ with appropriate values, since the 
reactions Eqs.~(\ref{eq:D++}) - (\ref{eq:D0}) still 
act to randomize the nucleon isospins with the modified 
probabilities determined by the detailed balance
condition.

Here, we emphasize that the large pion density in the 
hadronic medium is responsible for the validity of 
Eq.~(\ref{eq:gg}) in the final state.
In the hadronic medium, there are so many pions which 
can be regarded as a heat bath when the nucleon sector 
is concerned, while nucleons are so dilutely distributed 
that they do not feel other ones' existence.

So far, we have limited our attention to reactions mediated 
by $\Delta(1232)$.
Interactions of nucleons with other mesons, however, can 
also take place in the hadronic medium, while they are
much less dominant.
It is also possible that $\Delta$ interacts with
thermal pions to form another resonance before its decay 
\cite{Pang:1992sk}.
All these reactions with thermal particles, however, 
proceed with certain probabilities determined by the 
isospin SU(2) symmetry as long as they are caused by the 
strong interaction.
Each reaction of a nucleon thus makes its isospin random,
and act to realize the factorization Eq.~(\ref{eq:P}).

\subsubsection{Low beam-energy region}

The factorization Eq.~(\ref{eq:gg}) is fully established 
for the RHIC energy.
At very low beam energy, however, pions are not produced
enough and the duration of the hadron phase below $T_{\rm chem}$ 
becomes shorter.
Nucleons, therefore, will not undergo sufficient charge exchange 
reactions below $T_{\rm chem}$.
When the reactions hardly occur, the isospin correlations 
generated at the hadronization remain until the final state.
At low beam energy, also the density of the nucleon becomes 
comparable to that of pions, and pions can no longer be regarded 
as a heat bath to absorb isospin fluctuations of nucleons.
The requirements to justify the factorization Eq.~(\ref{eq:gg}), 
therefore, eventually breaks down as the beam energy is decreased.
This would happen when $T_{\rm chem} \lesssim m_\pi$, since 
the abundance of pions is responsible for all of the above 
conditions.
From the $\sqrt{s_{\rm NN}}$ dependence of the chemical freezeout 
line on the $T$-$\mu_{\rm B}$ plane \cite{Cleymans:1998fq},
the factorization Eq.~(\ref{eq:gg}) should be 
well-satisfied in the range of beam energy 
$\sqrt{s_{\rm NN}} \gtrsim 10$GeV.

\subsection{Strange baryons}
\label{sec:B:S}

So far, we have limited our attention to nucleons.
Since baryons in the final state in heavy ion collisions 
are dominated by nucleons, the nucleon number,
which is not a conserved charge, is qualitatively 
identified with the baryon one.
It is, however, important to recognize the difference 
between these two fluctuation observables 
especially in considering higher-order cumulants.
The difference predominantly comes 
from strange baryons $\Lambda$ and $\Sigma$.
In this subsection, we argue a practical method to 
include the effect of these degrees of freedom in our 
factorization formula.

Strange baryons produced in the hadronic medium decay via 
the weak or electromagnetic interaction outside the fireball. 
$\Lambda$ decays via the weak interaction into $p\pi^-$ and $n\pi^0$ 
with the branching ratio 
\begin{align}
P_{\Lambda \to p} : P_{\Lambda \to n} \simeq 16:9.
\label{eq:P_L}
\end{align}
On the other hand, branching ratio of $\Sigma^+$ is
\begin{align}
P_{\Sigma^+ \to p} : P_{\Sigma^+ \to n} \simeq 13:12,
\label{eq:P_Sp}
\end{align}
while $\Sigma^-$ always decays into $n\pi^-$.
$\Sigma^0$ decays into $\Lambda$ via the electromagnetic interaction 
and then decays with Eq.~(\ref{eq:P_L}) \cite{PDG}.
If the $\Lambda$ and $\Sigma$ multiplets are created with an
equal probability, the production ratio of $p$ and $n$ from
their decays is given by
\begin{align}
P_{\Lambda,\Sigma\to p}:P_{\Lambda,\Sigma\to n} \simeq 9:11.
\end{align}
Actually, because of the mass splitting between $\Lambda$ and
the $\Sigma$ triplets, $\delta m \simeq T_{\rm chem}/2$, the
production of the $\Sigma$ triplets are a bit suppressed 
compared to that of $\Lambda$. This makes the above ratio
even closer to even.
If one can assume that the correlations between 
strange baryons emitted from the fireball are negligible,
therefore, the number of
nucleons produced by these decays can be incorporated into 
$N_p$ and $N_n$ in Eq.~(\ref{eq:P}).
The nucleon number in Eq.~(\ref{eq:P}), then, is promoted 
to that of baryons.
The same argument holds also for
$\bar{\Lambda}$ and $\bar{\Sigma}$.

In short, by simply counting all protons observed by 
detectors in the event-by-event analysis, 
$N_{\rm N}$ and $N_{\bar{\rm N}}$ in Eq.~(\ref{eq:P}) are 
automatically promoted to the baryon and anti-baryon 
numbers, respectively.

\section{Relating baryon and proton number cumulants}
\label{sec:cumulant}

In this section, we focus on the cumulants of the baryon and 
proton numbers, and derive formulas to relate these 
cumulants on the basis of the factorization Eq.~(\ref{eq:P}).
With these relations the cumulants of the baryon 
number, which is a conserved charge, are calculated
from experimentally observed proton number ones.

In this section, we change the variables in the probability 
distribution function in Eq.~(\ref{eq:P}) as 
\begin{align}
{\cal P}(N_p,N_{\bar p};N_{\rm B},N_{\bar{\rm B}})
={\cal P}_N(N_p,N_n,N_{\bar p},N_{\bar n}),
\end{align}
where we have replaced the neutron numbers with the baryon ones, 
$N_{\rm B}=N_p+N_n$ and $N_{\bar{\rm B}}=N_{\bar p}+N_{\bar n}$.
It is understood that the prescription discussed 
in Sec.~\ref{sec:B:S} is adopted for $\Lambda$, $\Sigma$,
and their antiparticles.

\subsection{Probability distribution functions}
\label{sec:cumulant:P}

Before deriving formulas to relate the baryon and proton 
number cumulants, in this subsection we first remark 
that the distribution functions of these degrees of freedom 
satisfy a linear relation under 
the factorization Eq.~(\ref{eq:P}).
This relation explains why the baryon 
number cumulants can be represented by 
the proton number cumulants and vice versa.

Let us start with the final state
proton and anti-proton number distribution function,
Eq.~(\ref{eq:gg}),
\begin{align}
{\cal G}(N_p,N_{\bar p})
&= \sum_{N_{\rm B},N_{\bar{\rm B}}} 
{\cal P}(N_p,N_{\bar p};N_{\rm B},N_{\bar{\rm B}})
\nonumber \\
&= \sum_{N_{\rm B},N_{\bar{\rm B}}} {\cal F}(N_{\rm B},N_{\bar{\rm B}}) 
M( N_p,N_{\bar p} ; N_{\rm B},N_{\bar{\rm B}} )
\label{eq:G=FM}
\end{align}
with 
\begin{align}
M( N_p,N_{\bar p} ; N_B,N_{\bar{\rm B}} ) 
= B_r ( N_p;N_{\rm B} ) B_{\bar r} ( N_{\bar p} ; N_{\bar{\rm B}} ).
\label{eq:KBB}
\end{align}
Equation (\ref{eq:G=FM}) shows that the distribution functions 
${\cal G}(N_p,N_{\bar p})$ and 
${\cal F}(N_{\rm B},N_{\bar{\rm B}})$ satisfy a linear relation.
Since $M( N_p,N_{\bar p} ; N_{\rm B},N_{\bar{\rm B}} )$ has 
the inverse, $M^{-1}( N_B,N_{\bar B} ; N_p,N_{\bar p} )$, 
${\cal F}(N_{\rm B},N_{\bar{\rm B}})$ is given in terms of
${\cal G}(N_p,N_{\bar p})$ as 
\begin{align}
{\cal F}(N_{\rm B},N_{\bar{\rm B}})
= \sum_{N_p,N_{\bar p}} {\cal G}(N_p,N_{\bar p})
M^{-1}( N_{\rm B},N_{\bar{\rm B}} ; N_p,N_{\bar p} ).
\label{eq:F=GM}
\end{align}
The specific form of $M^{-1}( N_{\rm B},N_{\bar{\rm B}} ; N_p,N_{\bar p} )$
is easily obtained by using the fact that the matrix 
Eq.~(\ref{eq:KBB}) has a triangular structure, in the 
sense that $M( N_p,N_{\bar p} ; N_{\rm B},N_{\bar{\rm B}} )$ 
takes nonzero values only for $N_p \leq N_{\rm B}$ and 
$N_{\bar p} \leq N_{\bar{\rm B}}$.
Using Eq.~(\ref{eq:F=GM}), the baryon number distribution 
function ${\cal F}(N_{\rm B},N_{\bar{\rm B}})$ 
\cite{BraunMunzinger:2011dn} 
is in principle determined by ${\cal G}(N_p,N_{\bar p})$.
In practice, however, this analysis does not work 
efficiently since the elements of 
$M^{-1}( N_{\rm B},N_{\bar{\rm B}}; N_p,N_{\bar p} )$
are rapidly oscillating, which results in
large errorbars in ${\cal F}(N_{\rm B},N_{\bar{\rm B}})$ 
determined in this way.
In the following, instead of the distribution functions 
themselves, we concentrate on the cumulants of 
${\cal F}(N_{\rm B},N_{\bar{\rm B}})$ and 
${\cal G}(N_p,N_{\bar p})$.

\subsection{Generating functions and Cumulants}
\label{sec:cumulant:gf}

The moments and cumulants of a distribution function
are defined in terms of their generating functions.
The moment generating function for the proton and 
anti-proton numbers with the probability 
${\cal P}(N_p,N_{\bar p};N_{\rm B},N_{\bar{\rm B}})$
is given by 
\begin{align}
G( \theta,\bar\theta ) 
= \sum_{N_p,N_{\bar p},N_{\rm B},N_{\bar{\rm B}}} 
{\cal P} (N_p,N_{\bar p};N_{\rm B},N_{\bar{\rm B}})
e^{N_p\theta} e^{N_{\bar p}\bar\theta},
\label{eq:G}
\end{align}
and the corresponding cumulant generating function reads 
\begin{align}
K( \theta,\bar\theta ) = \log G( \theta,\bar\theta ) .
\label{eq:K}
\end{align}
Derivatives of Eq.~(\ref{eq:G}) give moments of 
${\cal P}(N_p,N_{\bar p};N_{\rm B},N_{\bar{\rm B}})$,
\begin{align}
\langle N_p^n N_{\bar p}^m \rangle
=  \left . \frac{ \partial^n }{ \partial \theta^n }
\frac{ \partial^m }{ \partial \bar\theta^m }
G( \theta,\bar\theta ) \right |_{\theta=\bar{\theta}=0} ,
\label{eq:moments}
\end{align}
as long as the sum in Eq.~(\ref{eq:G}) converges,
while cumulants of the proton and anti-proton numbers are 
defined with Eq.~(\ref{eq:K}) as 
\begin{align}
\langle (\delta N_p)^n (\delta N_{\bar p})^m \rangle_c
= \left . \frac{ \partial^n }{ \partial \theta^n }
\frac{ \partial^m }{ \partial \bar\theta^m }
K( \theta,\bar\theta ) \right |_{\theta=\bar{\theta}=0} .
\label{eq:cumulant}
\end{align}
The first-order cumulant is the expectation value of 
the operator
\begin{align}
\langle \delta N_p \rangle_c = \langle N_p \rangle, \quad
\langle \delta N_{\bar p} \rangle_c = \langle N_{\bar p} \rangle,
\end{align}
while the second- and third-order cumulants are moments of fluctuations,
such as,
\begin{align}
\langle \delta N_p \delta N_{\bar p}\rangle_c 
= \langle \delta N_p \delta N_{\bar p}\rangle,
\end{align}
and so forth, with $\delta N_X = N_X - \langle N_X \rangle$.

Substituting the explicit form of 
${\cal P}(N_p,N_{\bar p};N_{\rm B},N_{\bar{\rm B}})$ in 
Eq.~(\ref{eq:gg}) for $K(\theta,\bar\theta)$, one obtains 
\begin{align}
K( \theta,\bar\theta ) 
= \log \sum_{N_{\rm B},N_{\bar{\rm B}}} {\cal F}(N_{\rm B},N_{\bar{\rm B}})
\exp( k_{N_{\rm B},N_{\bar{\rm B}}}(\theta,\bar\theta) ),
\label{eq:Kk}
\end{align}
where
\begin{align}
\lefteqn{ k_{N_{\rm B},N_{\bar{\rm B}}}( \theta,\bar\theta ) }
\nonumber \\
&= \log \sum_{N_p} B_r(N_p;N_{\rm B}) e^{N_p\theta} +
\log \sum_{N_{\bar p}} B_{\bar r}(N_{\bar p};N_{\bar{\rm B}}) 
e^{N_{\bar p}\bar \theta},
\label{eq:k}
\end{align}
is the cumulant generating function for 
two independent binomial distribution functions.
With Eq.~(\ref{eq:k}), one easily finds that this function
satisfies $k_{N_{\rm B},N_{\bar{\rm B}}}(0,0)=0$ and 
\begin{align}
\frac{ \partial^n }{ \partial \theta^n } k_{N_{\rm B},N_{\bar{\rm B}}}( 0,0 ) 
&= \xi_n N_{\rm B},
\label{eq:k1}
\\
\frac{ \partial^m }{ \partial \bar\theta^m } k_{N_{\rm B},N_{\bar{\rm B}}}( 0,0 ) 
&= \bar\xi_m N_{\bar{\rm B}},
\label{eq:k2}
\\
\frac{ \partial^{n+m} }{ \partial \theta^n \partial \bar\theta^m } 
k_{N_{\rm B},N_{\bar{\rm B}}}( 0,0 ) &= 0,
\label{eq:k3}
\end{align}
for positive integers $n$ and $m$, with the cumulants of 
the binomial distribution function normalized by the total number
\begin{align}
\xi_1 &= r, \quad
\xi_2 = r(1-r), \quad
\xi_3 = r(1-r)(1-2r), 
\nonumber \\
\xi_4 &= r(1-r)(1-6r+6r^2), \quad \cdots,
\label{eq:xi}
\end{align}
and the same formulas for the anti-particle sector.
Imposing Eqs.~(\ref{eq:Kk}) - (\ref{eq:k3}) as the structure 
of $K(\theta,\bar\theta)$, cumulants of net proton and baryon 
numbers, $N_p^{\rm (net)} = N_p - N_{\bar p}$ and 
$N_{\rm B}^{\rm (net)} = N_{\rm B} - N_{\bar{\rm B}}$, respectively, 
are calculated to be
\begin{widetext}
\begin{align}
\langle N_p^{\rm (net)} \rangle
=& \langle \xi_1 N_{\rm B} - \bar\xi_1 N_{\bar{\rm B}} \rangle,
\label{eq:Np1x}
\\
\langle (\delta N_p^{\rm (net)})^2 \rangle
=& \langle ( \xi_1 \delta N_{\rm B} 
- \bar\xi_1 \delta N_{\bar{\rm B}} )^2 \rangle
+ \langle \xi_2 N_{\rm B} + \bar\xi_2 N_{\bar{\rm B}} \rangle,
\label{eq:Np2x}
\\
\langle (\delta N_p^{\rm (net)})^3 \rangle
=& \langle ( \xi_1 \delta N_{\rm B} 
- \bar\xi_1 \delta N_{\bar{\rm B}} )^3 \rangle
+ 3 \langle ( \xi_2 \delta N_{\rm B} + \bar\xi_2 \delta N_{\bar{\rm B}} )
( \xi_1 \delta N_{\rm B} - \bar\xi_1 \delta N_{\bar{\rm B}} ) \rangle
+ \langle \xi_3 N_{\rm B} - \bar\xi_3 N_{\bar{\rm B}} \rangle,
\label{eq:Np3x}
\\ 
\langle (\delta N_p^{\rm (net)})^4 \rangle_c
=& \langle ( \xi_1 \delta N_{\rm B} 
- \bar\xi_1 \delta N_{\bar{\rm B}} )^4 \rangle_c
+ 6 \langle ( \xi_2 \delta N_{\rm B} + \bar\xi_2 \delta N_{\bar{\rm B}} )
( \xi_1 \delta N_{\rm B} - \bar\xi_1 \delta N_{\bar{\rm B}} )^2 \rangle
\nonumber \\ 
&+ 3 \langle ( \xi_2 \delta N_{\rm B} 
+ \bar\xi_2 \delta N_{\bar{\rm B}} )^2 \rangle
+ 4 \langle ( \xi_3 \delta N_{\rm B} - \bar\xi_3 \delta N_{\bar{\rm B}} )
( \xi_1 \delta N_{\rm B} - \bar\xi_1 \delta N_{\bar{\rm B}} ) \rangle
+ \langle \xi_4 N_{\rm B} + \bar\xi_4 N_{\bar{\rm B}} \rangle,
\label{eq:Np4x}
\end{align}
and 
\begin{align}
\langle N_{\rm B}^{\rm (net)} \rangle
=& \left\langle \xi_1^{-1} N_p 
- \bar\xi_1^{-1} N_{\bar p} \right\rangle,
\label{eq:NB1x}
\\
\left\langle (\delta N_{\rm B}^{\rm (net)})^2 \right\rangle
=& 
\left\langle \left( \xi_1^{-1} \delta N_p 
- \bar{\xi}_1^{-1} \delta N_{\bar p} \right)^2 \right\rangle
- \left\langle \xi_2 \xi_1^{-3} \delta N_p 
+ \bar\xi_2 \bar\xi_1^{-3} \delta N_{\bar p} \right\rangle,
\label{eq:NB2x}
\\
\left\langle (\delta N_{\rm B}^{\rm (net)} )^3 \right\rangle
=& \left\langle \left( \xi_1^{-1} \delta N_p 
- \bar{\xi}_1^{-1} \delta N_{\bar p} \right)^3 \right\rangle
-3 \left\langle \left( \xi_2 \xi_1^{-3} \delta N_p 
+ \bar\xi_2 \bar\xi_1^{-3} \delta N_{\bar p} \right)
\left( \xi_1^{-1} \delta N_p - \bar{\xi}_1^{-1} \delta N_{\bar p} 
\right) \right\rangle
\nonumber 
\\
&+ \left\langle \frac{ 3\xi_2^2 - \xi_1 \xi_3 }{ \xi_1^5 } N_p
- \frac{ 3\bar\xi_2^2 - \bar\xi_1 \bar\xi_3 }{ \bar\xi_1^5 } 
N_{\bar p} \right\rangle,
\label{eq:NB3x}
\\
\left\langle ( \delta N_{\rm B}^{\rm (net)} )^4 \right\rangle_c
=& \left\langle \left( \xi_1^{-1} \delta N_p 
- \bar{\xi}_1^{-1} \delta N_{\bar p} \right)^4 \right\rangle_c
-6 \left\langle \left( \xi_2 \xi_1^{-3} \delta N_p 
+ \bar\xi_2 \bar\xi_1^{-3} \delta N_{\bar p} \right)
\left( \xi_1^{-1} \delta N_p - \bar{\xi}_1^{-1} \delta N_{\bar p} 
\right) \right\rangle
\nonumber 
\\
& + 12 \left\langle \left( \xi_2^2 \xi_1^{-5} \delta N_p 
- \bar\xi_2^2 \bar\xi_1^{-5} \delta N_{\bar p} \right)
\left( \xi_1^{-1} \delta N_p - \bar{\xi}_1^{-1} \delta N_{\bar p} 
\right) \right\rangle
+3 \left\langle \left( \xi_2 \xi_1^{-3} \delta N_p 
+ \bar\xi_2 \bar\xi_1^{-3} \delta N_{\bar p} \right)^2 \right\rangle
\nonumber
\\
& - 4 \left\langle \left( \xi_3 \xi_1^{-4} \delta N_p
- \bar\xi_3 \bar\xi_1^{-4} \delta N_{\bar p} \right) 
\left( \xi_1^{-1} \delta N_p - \bar{\xi}_1^{-1} \delta N_{\bar p} 
\right) \right\rangle
\nonumber
\\
& - \left\langle 
\frac{ 15\xi_2^3 - 10 \xi_1 \xi_2 \xi_3 + \xi_1^2 \xi_4 }{ \xi_1^7 } N_p
- \frac{ 15\bar\xi_2^3 - 10 \bar\xi_1 \bar\xi_2 \bar\xi_3 
+ \bar\xi_1^2 \bar\xi_4 }{ \bar\xi_1^7 } N_{\bar p}
\right\rangle.
\label{eq:NB4x}
\end{align}
\end{widetext}
A detailed description of the procedure to obtain these results
is given in Appendix~\ref{app:cumulant}.
We emphasize that no explicit form of ${\cal F}(N_{\rm B},N_{\bar{\rm B}})$ 
is assumed in deriving these results.
Moreover, in Appendix~\ref{app:cumulant} 
we only use Eq.~(\ref{eq:Kk}) for the structure of 
$K(\theta,\bar\theta)$ and Eqs.~(\ref{eq:k1}) - (\ref{eq:k3}) 
for properties of $k_{N_{\rm B},N_{\bar{\rm B}}}(\theta,\bar\theta)$
to derive Eqs.~(\ref{eq:Np1x}) - (\ref{eq:NB4x}).
Therefore, these results 
hold for any distribution functions satisfying 
these conditions with the appropriate choice for the 
values of $\xi_i$ and $\bar\xi_i$.

\subsection{Isospin symmetric case}

In hot medium produced by heavy ion collisions,
(anti-)proton and (anti-)neutron number densities are 
in general different because of the isospin asymmetry 
of colliding heavy nuclei.
In relativistic heavy ion collisions at sufficiently
large $\sqrt{s_{\rm NN}}$ and small impact parameters, however, 
the isospin density is negligibly small
because a large number of particles having nonzero isospin 
charges (mainly pions) are created and most of the initial
isospin density is absorbed by these degrees of freedom
(see, Appendix~\ref{app:isospin}).
When the isospin density vanishes,
$r$ and $\bar{r}$ are to be set at $1/2$ in the binomial 
distribution functions in Eq.~(\ref{eq:P}).
Substituting 
\begin{align}
\xi_1 = \frac12 , \quad \xi_2 = \frac14, \quad
\xi_3 = 0 , \quad \xi_4 = -\frac18,
\label{eq:xi1/2}
\end{align}
into Eqs.~(\ref{eq:Np1x}) - (\ref{eq:NB4x}), 
which are obtained by putting $r=1/2$ in Eq.~(\ref{eq:xi}),
one obtains 
\begin{widetext}
\begin{align}
\langle N_p^{\rm (net)} \rangle
=& \frac12 \langle N_{\rm B}^{\rm (net)} \rangle,
\label{eq:Np1}
\\
\langle (\delta N_p^{\rm (net)})^2 \rangle
=& \frac14 \langle (\delta N_{\rm B}^{\rm (net)})^2 \rangle
+ \frac14 \langle N_{\rm B}^{\rm (tot)} \rangle,
\label{eq:Np2}
\\
\langle (\delta N_p^{\rm (net)})^3 \rangle
=& \frac18 \langle (\delta N_{\rm B}^{\rm (net)})^3 \rangle
+ \frac38 \langle \delta N_{\rm B}^{\rm (net)} \delta N_{\rm B}^{\rm (tot)} \rangle,
\label{eq:Np3}
\\
\langle (\delta N_p^{\rm (net)})^4 \rangle_c
=& \frac1{16} \langle (\delta N_{\rm B}^{\rm (net)})^4 \rangle_c
+ \frac38 \langle (\delta N_{\rm B}^{\rm (net)})^2 \delta N_{\rm B}^{\rm (tot)} \rangle
+ \frac3{16} \langle (\delta N_{\rm B}^{\rm (tot)})^2 \rangle
- \frac18 \langle N_{\rm B}^{\rm (tot)} \rangle,
\label{eq:Np4}
\end{align}
and 
\begin{align}
\langle N_{\rm B}^{\rm (net)} \rangle
=& 2 \langle N_p^{\rm (net)} \rangle ,
\label{eq:NB1}
\\
\langle (\delta N_{\rm B}^{\rm (net)})^2 \rangle
=& 4 \langle (\delta N_p^{\rm (net)})^2 \rangle
-2 \langle N_p^{\rm (tot)} \rangle ,
\label{eq:NB2}
\\
\langle (\delta N_{\rm B}^{\rm (net)})^3 \rangle
=& 8 \langle (\delta N_p^{\rm (net)})^3 \rangle
-12 \langle \delta N_p^{\rm (net)} \delta N_p^{\rm (tot)} \rangle
+6 \langle N_p^{\rm (net)} \rangle ,
\label{eq:NB3}
\\
\langle (\delta N_{\rm B}^{\rm (net)})^4 \rangle_c
=& 16 \langle (\delta N_p^{\rm (net)})^4 \rangle_c
-48 \langle (\delta N_p^{\rm (net)})^2 \delta N_p^{\rm (tot)} \rangle
+ 48 \langle (\delta N_p^{\rm (net)})^2 \rangle
+ 12 \langle (\delta N_p^{\rm (tot)})^2 \rangle
- 26 \langle N_p^{\rm (tot)} \rangle ,
\label{eq:NB4}
\end{align}
\end{widetext}
which are the results given in Ref.~\cite{Kitazawa:2011wh}.
Here a note is in order about the terms on RHSs of
Eqs.~(\ref{eq:NB2})-(\ref{eq:NB4}). Each term on RHS of
these equations is not necessarily uncorrelated with
each other. In particular, generally
${\cal F}(N_{\rm B},N_{\rm \bar{B}})$ is not separable,
i.e., it cannot be written as
${\cal F}(N_{\rm B},N_{\rm \bar{B}}) = f(N_{\rm B})g(N_{\rm \bar{B}})$.
If there is such correlation, the statistical fluctuations of
these terms are not independent but mutually correlated.
Thus, an appropriate care need to be taken in estimating
the statistical error for LHSs of 
Eqs.~(\ref{eq:NB2})-(\ref{eq:NB4}).

\subsection{Effect of nonzero isospin density}
\label{sec:cumulant:nid}

As the collision energy is lowered, the effect of nonzero
isospin density eventually gives rise to non-negligible
contribution to the above relations.
To investigate this effect, we first assume that the 
isospins of nucleons, anti-nucleons, and pions in the 
final state are in chemical equilibrium,
as is indicated by the fast N$\pi$ reactions discussed 
in the previous Section.
Because the nucleon distribution is well approximated
by the Boltzmann distribution, 
the numbers of (anti-)protons and (anti-)neutrons 
in the final state are given with the isospin 
chemical potential $\mu_{\rm I}$ and the temperature $T$ as 
\begin{align}
& \langle N_p \rangle = C e^{\mu_{\rm I}/(2T)}, \quad 
\langle N_{\bar p} \rangle = D e^{-\mu_{\rm I}/(2T)}, 
\nonumber \\
& \langle N_n \rangle = C e^{-\mu_{\rm I}/(2T)}, \quad 
\langle N_{\bar n} \rangle = D e^{\mu_{\rm I}/(2T)},
\end{align}
where $C$ and $D$ are constants determined by 
the chemical freezeout condition such as
the volume of the system, the rapidity coverage, and so on.
These relations lead to 
\begin{align}
\frac{ \langle N_p \rangle }{ \langle N_n \rangle }
= \frac{ \langle N_{\bar n} \rangle }{ \langle N_{\bar p} \rangle }
= e^{\mu_{\rm I}/T},
\end{align}
and thereby $r=1-\bar{r}$.
One thus can parametrize $r$ and $\bar{r}$ as 
\begin{align}
r = \frac12 - \alpha , \quad \bar{r} = \frac12 + \alpha,
\label{eq:ralpha}
\end{align}
with the negative isospin density per nucleon
\begin{align}
\alpha 
= \frac12 \cdot \frac{ \langle N_n\rangle - \langle N_p \rangle  }
{ \langle N_n \rangle + \langle N_p \rangle  }
= \frac12 \cdot 
\frac{ 1 - e^{\mu_{\rm I}/T} }{ 1 + e^{\mu_{\rm I}/T} } .
\label{eq:alpha}
\end{align}
$\alpha$ assumes a positive value in heavy ion collisions.

When the value of $\alpha$ is small, $\alpha\ll1$,
the effects of nonzero isospin density on Eqs.~(\ref{eq:NB1x}) 
- (\ref{eq:NB4x}) are well described by the 
Taylor series with respect to $\alpha$.
Substituting Eq.~(\ref{eq:ralpha}) in these equations,
up to the first order in $\alpha$ 
Eqs.~(\ref{eq:NB1}) - (\ref{eq:NB4}) become 
\begin{widetext}
\begin{align}
\langle N_{\rm B}^{\rm (net)} \rangle
=& 2 \langle N_p^{\rm (net)} \rangle
+ 4\alpha \langle N_p^{\rm (tot)} \rangle + O(\alpha^2),
\label{eq:NB1I}
\\
\langle (\delta N_{\rm B}^{\rm (net)})^2 \rangle
=& 4 \langle (\delta N_p^{\rm (net)})^2 \rangle
+ 2 \langle N_p^{\rm (tot)} \rangle
+ 4 \alpha \left( 
4 \langle \delta N_p^{\rm (net)}\delta N_p^{\rm (tot)} \rangle
-3 \langle N_p^{\rm (net)} \rangle \right) + O(\alpha^2),
\label{eq:NB2I}
\\
\langle (\delta N_{\rm B}^{\rm (net)})^3 \rangle
=& 8 \langle (\delta N_p^{\rm (net)})^3 \rangle
- 12 \langle \delta N_p^{\rm (net)} \delta N_p^{\rm (tot)} \rangle
+ 6 \langle N_p^{\rm (net)} \rangle
\nonumber \\
& + 4 \alpha \left( 
12 \langle (\delta N_p^{\rm (net)})^2 \delta N_p^{\rm (tot)} \rangle
- 18 \langle (\delta N_p^{\rm (net)})^2 \rangle
- 6 \langle (\delta N_p^{\rm (toe)})^2 \rangle
+ 13 \langle N_p^{\rm (tot)} \rangle \right) + O(\alpha^2),
\label{eq:NB3I}
\\
\langle (\delta N_{\rm B}^{\rm (net)})^4 \rangle_c
=& 16 \langle (\delta N_p^{\rm (net)})^4 \rangle_c
-48 \langle (\delta N_p^{\rm (net)})^2 \delta N_p^{\rm (tot)} \rangle
+ 48 \langle (\delta N_p^{\rm (net)})^2 \rangle
+ 12 \langle (\delta N_p^{\rm (tot)})^2 \rangle
- 26 \langle N_p^{\rm (tot)} \rangle 
\nonumber \\
& + 4 \alpha \left( 
32 \langle (\delta N_p^{\rm (net)})^3 \delta N_p^{\rm (tot)} \rangle_c
- 72 \langle (\delta N_p^{\rm (net)})^3 \rangle
- 48 \langle \delta N_p^{\rm (net)} (\delta N_p^{\rm (tot)})^2 \rangle
+ 164 \langle \delta N_p^{\rm (net)} \delta N_p^{\rm (tot)} \rangle
\right.
\nonumber \\
& \left.
- 75 \langle N_p^{\rm (net)} \rangle \right) + O(\alpha^2).
\label{eq:NB4I}
\end{align}
\end{widetext}

Next, let us estimate the value of $\alpha$ in relativistic heavy ion
collisions.
Under the chemical equilibrium condition, the ratio of 
the charged pion numbers, $\langle N_{\pi^+}\rangle$ and 
$\langle N_{\pi^-}\rangle$, having isospin charges $\pm1$, is given by
\begin{align}
\frac{ \langle N_{\pi^-} \rangle}{ \langle N_{\pi^+} \rangle} 
\simeq e^{-2\mu_{\rm I}/T},
\label{eq:n_pi_pm}
\end{align}
where we have adopted Boltzmann statistics for pions, 
since the effect of Bose-Einstein correlation on the pion 
density is about $10\%$ for $T_{\rm chem} = m_\pi$
and does not affect our qualitative conclusion.
The experimental result for 
$\langle N_{\pi^-} \rangle / \langle N_{\pi^+} \rangle$ in the final 
state is almost unity for high energy collisions in 
accordance with the approximate isospin symmetry.
Substituting Eq.~(\ref{eq:n_pi_pm}) in Eq.~(\ref{eq:alpha})
and using $\langle N_{\pi^-} \rangle / \langle N_{\pi^+} \rangle - 1 \ll 1$,
one obtains 
\begin{align}
\alpha \simeq \frac18 \left( 
\frac{ \langle N_{\pi^-} \rangle}{ \langle N_{\pi^+} \rangle} - 1 \right).
\label{eq:alpha-pi}
\end{align}
The value of $\alpha$, as well as 
$\langle N_{\pi^-} \rangle / \langle N_{\pi^+} \rangle - 1$,
grows as $\sqrt{s_{\rm NN}}$ is lowered.
In order to see how these parameters become non-negligible
for small $\sqrt{s_{\rm NN}}$, we focus on the $40$GeV collision 
at the SPS ($\sqrt{s_{\rm NN}}\simeq9$GeV).
For this collision, the experimental value of
$\langle N_{\pi^-} \rangle / \langle N_{\pi^+} \rangle$ is $1.05\pm0.05$ 
\cite{BraunMunzinger:2003zd}.
Substituting the worst value within $1\sigma$,
$\langle N_{\pi^-} \rangle / \langle N_{\pi^+} \rangle = 1.1$, 
in Eq.~(\ref{eq:alpha-pi}), 
one obtains $\alpha \simeq 1/80$.
On the other hand, below the top SPS energy the production 
of anti-nucleons is well suppressed and 
one can replace all $\delta N_p^{\rm (net)}$ and 
$\delta N_p^{\rm (tot)}$ in Eqs.~(\ref{eq:NB1I}) - (\ref{eq:NB4I}) 
with $\delta N_p$ to a good approximation.
Equation (\ref{eq:NB4I}), for example, then becomes
\begin{align}
\lefteqn{ \langle (\delta N_{\rm B}^{\rm (net)})^4 \rangle_c 
\simeq 16 (1+8\alpha)\langle (\delta N_p)^4 \rangle_c}
\nonumber \\
&-48 (1+10\alpha)\langle (\delta N_p)^3\rangle
+ 60(1+10.1\alpha) \langle (\delta N_p)^2 \rangle
\nonumber \\
&- 26(1+11.5\alpha) \langle N_p \rangle.
\label{eq:NB4alpha}
\end{align}
This result shows that for $\alpha = 1/80$ the corrections 
of nonzero isospin density to Eqs.~(\ref{eq:NB1}) - (\ref{eq:NB4}) 
are less than $10\%$ in magnitude.
The effect is smaller in relations for the lower-order 
cumulants, Eqs.~(\ref{eq:NB1I}) - (\ref{eq:NB3I}),
and formulas for proton number cumulants, Eqs.~(\ref{eq:Np1})
- (\ref{eq:Np4}).

With these results, one can conclude that the formulas 
for the isospin symmetric case, Eqs.~(\ref{eq:NB1}) 
- (\ref{eq:NB4}), can safely be used to the 
analysis of the baryon number cumulants 
for $\sqrt{s_{\rm NN}}\simeq 9{\rm GeV}$ with a precision 
of less than $10\%$.
Because the production of isospin charged particles
increases as $\sqrt{s_{\rm NN}}$ goes up, the value of $\alpha$,
and hence the effect of nonzero isospin density on 
Eqs.~(\ref{eq:NB1}) - (\ref{eq:NB4}) are more suppressed for 
higher energy collisions.

As $\sqrt{s_{\rm NN}}$ is lowered, the value of 
$\alpha$ grows and eventually approaches the one 
in the colliding heavy nuclei, $\alpha_A\simeq0.1$.
For $\alpha\simeq0.1$, the first-order correction in 
Eq.~(\ref{eq:NB4alpha}) is comparable with the zeroth-order
one.
Relations for the isospin symmetric case, 
Eqs.~(\ref{eq:NB1}) - (\ref{eq:NB4}), therefore, 
are no longer applicable.
For such collision energies, however, conditions required 
for the factorization Eq.~(\ref{eq:P}) themselves
break down as discussed in Sec.~\ref{sec:B:HIC}.

Before closing this subsection, we recapitulate that the suppression 
of the isospin density in the nucleon sector,
and hence $\alpha$, in the final 
state is caused by the production of the large number of 
particles having isospin charges, especially charged pions.
In Appendix~\ref{app:isospin}, we present an analysis for this effect.

\section{Discussions}
\label{sec:discussion}

\subsection{Recent experimental results on proton number cumulants}
\label{sec:discussion:exp}

As emphasized in the previous sections,
the cumulants of the proton and baryon numbers are in general
different.
One, therefore, has to be careful when comparing theoretical 
predictions on baryon number cumulants with experimental 
proton number ones.
In this subsection, we show that the deviation from the 
thermal distribution in baryon number cumulants becomes 
difficult to measure in proton number cumulants 
using relations obtained in the previous section with 
some additional assumptions.

In general, it is possible that, 
while the net baryon number fluctuations in the final state 
have a considerable deviation from the grand canonical ones 
reflecting the hysteresis of fireballs and/or the global 
charge conservation, baryon and anti-baryon 
numbers separately follow the thermal (Boltzmann) distributions.
For example, if the net baryon number fluctuations above $T_{\rm c}$ survive
until the final state, the net baryon number fluctuations remain
small compared to the thermal ones in the hadronic medium,
while baryon and anti-baryon number fluctuations separately 
follow the thermal one.
Generally, cumulants of net numbers cannot take arbitrary values;
for instance, the second-order cumulant is 
constrained by the Cauchy-Schwartz inequality:
\begin{align}
\lefteqn{
\left( \sqrt{\langle (\delta N_{\rm B})^2 \rangle }
- \sqrt{\langle (\delta N_{\bar{\rm B}})^2 \rangle } \right)^2 }
\nonumber \\
&\le \langle (\delta N_{\rm B}^{\rm (net)} )^2 \rangle
\le \left( \sqrt{\langle (\delta N_{\rm B})^2 \rangle }
+ \sqrt{\langle (\delta N_{\bar{\rm B}})^2 \rangle } \right)^2.
\label{eq:CS}
\end{align}
The values of net baryon number cumulants satisfying these 
constraints are not forbidden.
Suppose that, as an extreme case,
the net baryon number fluctuations completely vanish and
the left equality in Eq. (\ref{eq:CS}) is realized. 
A baryon and anti-baryon distribution function 
\begin{align}
{\cal F}(N_{\rm B},N_{\bar{\rm B}}) 
= P_\lambda (N_{\rm B}) \delta_{N_{\rm B},N_{\bar{\rm B}}},
\end{align}
which is a constrained baryon and anti-baryon number 
distribution following the canonical distribution,
constitutes such an example.
The distribution function 
${\cal F}(N_{\rm B},N_{\bar{\rm B}}) $ for free gas in 
the grand canonical ensemble, i.e.,
an unconstrained case,
on the other hand, is given by Eq.~(\ref{eq:F_HG}).

Now, let us consider the difference between the net baryon and 
net proton number cumulants when the baryon and anti-baryon number
distributions follow Boltzmann statistics while the net baryon
number does not.
Because of the Boltzmann nature of $N_{\rm B}$ 
and $N_{\bar{\rm B}}$, distributions of $N_p$ and $N_{\bar p}$ 
are also poissonian from Eq.~(\ref{eq:P}).
Thus, cumulants of the baryon and proton numbers satisfy
\begin{align}
\lefteqn{
\langle N_{\rm B} \rangle
= \langle (\delta N_{\rm B})^2 \rangle
= \langle (\delta N_{\rm B})^3 \rangle
= 2 \langle N_p \rangle_{\rm HG} 
}
\nonumber \\
&= 2 \langle ( \delta N_p )^2 \rangle_{\rm HG}
= 2 \langle ( \delta N_p )^3 \rangle_{\rm HG}
= 2 \langle ( \delta N_p )^4 \rangle_{c,\rm HG},
\label{eq:HG}
\end{align}
and the same for the anti-baryon and anti-proton numbers, where 
$\langle \cdot \rangle_{\rm HG}$ is the expectation value for 
free hadron gas (HG) composed of mesons and nucleons at 
$T_{\rm chem}$, i.e., a simplified version of the HRG 
model \cite{Karsch:2010ck}.
The factors two in front of the proton number cumulants 
in Eq.~(\ref{eq:HG}) are understood from
Eq.~(\ref{eq:PP=PB}).

Using Eq.~(\ref{eq:HG}), the second terms in 
Eqs.~(\ref{eq:Np2}) and (\ref{eq:Np3}) are transformed as
\begin{align}
\langle N_{\rm B}^{\rm (tot)} \rangle
&= 2 \langle (\delta N_p)^2 + (\delta N_{\bar p})^2 \rangle_{\rm HG}
\nonumber \\
&= 2 \langle (\delta N_p^{\rm (net)})^2 \rangle_{\rm HG},
\label{eq:HG1}
\\
\langle \delta N_{\rm B}^{\rm (net)} \delta N_{\rm B}^{\rm (tot)} \rangle
&= \langle (\delta N_{\rm B})^2 - (\delta N_{\bar{\rm B}})^2 \rangle
\nonumber \\
&= 2 \langle (\delta N_p)^3 - (\delta N_{\bar p})^3 \rangle_{\rm HG}
\nonumber \\
&= 2 \langle (\delta N_p^{\rm (net)})^3 \rangle_{\rm HG},
\label{eq:HG2}
\end{align}
where in the last equalities we have used 
the fact that the proton and anti-proton numbers do not have 
correlations in the free gas, i.e.,
$ \langle \delta N_p \delta N_{\bar p} \rangle_{\rm HG}
= \langle (\delta N_p)^2 \delta N_{\bar p} \rangle_{\rm HG}
= \langle \delta N_p (\delta N_{\bar p})^2 \rangle_{\rm HG} =0 $.
Substituting Eqs.~(\ref{eq:HG1}) and (\ref{eq:HG2}) in
Eqs.~(\ref{eq:Np2}) and (\ref{eq:Np3}), respectively, one 
obtains
\begin{align}
\langle (\delta N_p^{\rm (net)} )^2 \rangle
&= \frac14 \langle (\delta N_{\rm B}^{\rm (net)})^2 \rangle
+ \frac12 \langle (\delta N_p^{\rm (net)} )^2 \rangle_{\rm HG},
\label{eq:Np2HG}
\\
\langle (\delta N_p^{\rm (net)} )^3 \rangle
&= \frac18 \langle (\delta N_{\rm B}^{\rm (net)})^3 \rangle
+ \frac34 \langle (\delta N_p^{\rm (net)} )^3 \rangle_{\rm HG}.
\label{eq:Np3HG}
\end{align}
These results show that the second terms on the RHSs,
which come from the binomial distributions of the nucleon 
isospin, have significant contribution to the cumulants of
the proton number, and the contribution of the net baryon 
number cumulants,
$\langle (\delta N_{\rm B}^{\rm (net)} )^n \rangle$,
are relatively suppressed.
Since the second terms give the thermal fluctuations, 
these results show that the deviation of
$\langle (\delta N_{\rm B}^{\rm (net)} )^n \rangle$
from the thermal value is hard to be seen in the proton 
number cumulants.
Although one cannot transform the fourth-order relation
Eq.~(\ref{eq:Np4}) to a simple form as in Eqs.~(\ref{eq:Np2HG})
and (\ref{eq:Np3HG}), from the factor $1/16$ in front of 
$\langle (\delta N_B^{\rm (net)} )^4 \rangle_c$ in 
Eq.~(\ref{eq:Np4}) it is obvious that the direct contribution 
of this term to experimentally 
measured $\langle (\delta N_p^{\rm (net)} )^4 \rangle_c$ 
is more suppressed compared to the lower-order cumulants,
and that its experimental confirmation is more difficult.
These analyses strongly indicate that, even if 
the baryon number cumulants have considerable deviation
from the thermal values, they are obscured in the 
experimentally measured proton number cumulants
due to the redistribution in isospin space.
Such a tendency seems to become more 
prominent for higher-order cumulants.
It is known that higher-order cumulants of the baryon number 
have large critical exponents and thus can have significant 
enhancement in the vicinity of the 
critical point \cite{Stephanov:2008qz}.
The above result, however, indicates that such enhancement
is suppressed by a factor $1/2^n$ and difficult to measure 
in experiments in proton number cumulants.
The analysis of the baryon number cumulants with 
Eqs.~(\ref{eq:NB1}) - (\ref{eq:NB4}) enables to remove 
the thermal contribution in the proton number cumulants and 
makes the direct experimental observation of signals in 
$\langle (\delta N_p^{\rm (net)} )^n \rangle_c$ possible.

The $\sqrt{s_{\rm NN}}$ dependences of proton number cumulants
are recently measured by STAR collaboration at RHIC
\cite{Aggarwal:2010wy,Mohanty:2011nm}.
The experimental result shows that ratios between 
net proton number cumulants follow the prediction of 
the HRG model within about $10\%$ precision.
We, however, emphasize that one should not conclude 
from this result that baryon number cumulants also follow 
the prediction of the HRG model within $10\%$ precision.
As demonstrated above, the binomial nature of isospin 
distribution makes proton number cumulants close to
the ones in the HRG model.
In this sense, it is interesting that the experimental 
results for skewness and kurtosis nevertheless have 
small but significant deviations from the HRG 
predictions \cite{Mohanty:2011nm}.
The deviation, for example, in skewness, can be a 
consequence of $\langle (\delta N_{\rm B}^{\rm (net)})^3 \rangle$ 
in Eq.~(\ref{eq:Np3HG}), which possibly reflects
the properties of the matter in the early stage.

A remark on Eqs.~(\ref{eq:Np2HG}) and (\ref{eq:Np3HG}) 
is in order.
These formulas are obtained with the assumption that baryon 
and anti-baryon number distributions are poissonian, 
while the net baryon number is not.
When one further assumes that the net baryon number cumulants
also follow the thermal distribution in these results, 
these formulas simply reproduce the free gas result 
\begin{align}
\langle (\delta N_{\rm B}^{\rm (net)})^n \rangle_c
= 2\langle (\delta N_p^{\rm (net)})^n \rangle_{c,\rm HG}
\end{align}
as they should do.
This is easily checked by substituting 
$\langle (\delta N_p^{\rm (net)} )^n \rangle
= \langle (\delta N_p^{\rm (net)} )^n \rangle_{\rm HG}$
in Eqs.~(\ref{eq:Np2HG}) and (\ref{eq:Np3HG}).

In this subsection, we considered the experimental 
results on proton number cumulants using the results 
in Sec.~\ref{sec:cumulant}.
More direct application of these formulas, i.e., 
to determine baryon number
cumulants from experimental results on proton
number cumulants with 
Eqs.~(\ref{eq:NB1}) - (\ref{eq:NB4}), is to be done.
The baryon number cumulants obtained in this way 
are to be compared with various theoretical predictions.

\subsection{Efficiency and acceptance corrections}
\label{sec:discussion:B}

So far, we have considered the reconstruction 
of the missing information for the neutron number in 
experiments using Eq.~(\ref{eq:P}).
It is possible to extend this argument to 
infer different information on the event-by-event 
analysis.

An example is the evaluation of the effect
of efficiency and acceptance of detectors.
The experimental detectors usually do not have
$2\pi$ acceptance. Moreover, protons entering a detector 
are identified with some efficiency.
If one can assume that protons (anti-protons) in the final state is 
detected by the detector with a fixed probability
$\sigma$ ($\bar{\sigma}$) independent of momentum, 
multiplicity, and so on,
and the efficiency for each particle does not have correlations,
the distribution function ${\cal G}^{\rm (obs)}(
N_p^{\rm (obs)},N_{\bar p}^{\rm (obs)})$ for 
the observed proton and anti-proton numbers,
$N_p^{\rm (obs)}$ and $N_{\bar p}^{\rm (obs)}$,
respectively, are related to the one for all particles 
entering the detector, $N_p$ and $N_{\bar p}$, as 
\begin{align}
\lefteqn{
{\cal G}^{\rm (obs)}(N_p^{\rm (obs)},N_{\bar p}^{\rm (obs)})
}
\nonumber \\
&= \sum_{N_p,N_{\bar p}} {\cal G}(N_p,N_{\bar p})
B_\sigma ( N_p^{\rm (obs)};N_p ) 
B_{\bar{\sigma}}(N_{\bar p}^{\rm (obs)};N_{\bar p}),
\end{align}
or substituting this result in Eq.~(\ref{eq:P}) and 
using the property of the binomial distribution
one obtains
\begin{align}
\lefteqn{
{\cal P}^{\rm (obs)}(N_p^{\rm (obs)},N_{\bar p}^{\rm (obs)};
N_{\rm B},N_{\bar{\rm B}})
}
\nonumber \\
&= \sum_{N_p,N_{\bar p}} {\cal F}(N_{\rm B},N_{\bar{\rm B}})
B_{\sigma/2} ( N_p^{\rm (obs)};N_p ) 
B_{\bar{\sigma}/2} (N_{\bar p}^{\rm (obs)};N_{\bar p}).
\label{eq:Pobs}
\end{align}

Eq.~(\ref{eq:Pobs}) indicates that when the 
deviations of $\sigma$ and $\bar{\sigma}$ 
from the unity become large,
they affect cumulants with different orders
differently.
The effect of efficiency, therefore, cannot be canceled 
out by taking the ratio between cumulants.
In particular, as $\sigma$ and $\bar{\sigma}$ become smaller, 
${\cal G}^{\rm (obs)}(N_p^{\rm (obs)},N_{\bar p}^{\rm (obs)})$
approach the product of independent Poisson distributions
irrespective of the form of ${\cal F}(N_{\rm B},N_{\bar{\rm B}})$.
This would be another reason of the present 
experimental results on proton number cumulants 
\cite{Mohanty:2011nm}, 
which is consistent with the HRG model.

Other experimental artifacts which have not taken into account yet
in experimental analyses are background and misidentified protons.
In particular, according to Ref.~{\cite{Abelev:2008ab}},
the contamination from knockout protons is not negligible.
By their nature, they give poissonian contribution and
make observed proton number cumulants approach the poissonian values.
Indeed, the HIJING + GEANT simulation in Ref.~\cite{Aggarwal:2010wy}
shows that these effects are considerable.

\section{Summary}

The most important results of the present paper
is summarized in 
Eqs.~(\ref{eq:Np1}) - (\ref{eq:Np4}) and
Eqs.~(\ref{eq:NB1}) - (\ref{eq:NB4}), 
which are formulas relating baryon and proton number 
cumulants in the final state in heavy ion collisions.
The baryon number cumulants are a conserved charge, and 
one of the fluctuation observables which is most widely 
analyzed by theoretical studies.
Our results enable to determine the baryon number 
cumulants with experimental results in heavy ion collisions, 
and hence make the direct comparison between theoretical 
predictions and experiments possible.
Such a comparison will provide significant information 
on the QCD phase diagram.
The results Eqs.~(\ref{eq:Np1}) - (\ref{eq:NB4}) are obtained 
on the basis of the binomial nature of the nucleon and 
anti-nucleon number distributions in isospin space, which is 
justified for $\sqrt{s_{\rm NN}}\gtrsim 10{\rm GeV}$.
Although these results are obtained for isospin symmetric
medium, the effect of nonzero isospin density in relativistic
heavy ion collisions is well suppressed in this energy range
because of the abundance of the created pions.

The authors thank stimulating discussions at the workshop
``Fluctuations, Correlations and RHIC Low Energy Runs'' 
held at the Brookhaven National Laboratory, U.S.A., 
Oct 3rd through 5th, 2011.
This work is supported in part by Grants-in-Aid for 
Scientific Research by Monbu-Kagakusyo of Japan 
(No.~21740182 and 23540307).

\appendix

\section{Baryon and proton number cumulants}
\label{app:cumulant}

In this Appendix, we derive Eqs.~(\ref{eq:Np1x}) - 
(\ref{eq:NB4x}).
To obtain these relations, we start from 
the cumulant generating function Eq.~(\ref{eq:Kk}),
\begin{align}
K( \theta,\bar\theta ) 
= \log \sum_F
\exp\left[ k(\theta,\bar\theta) \right],
\label{eq:ap:Kk}
\end{align}
where $\sum_F$ is a shorthand notation for 
$\sum_{N_{\rm B},N_{\bar{\rm B}}} {\cal F}(N_{\rm B},N_{\bar{\rm B}})$.
In this Appendix, we also suppress the subscript in 
$k_{N_{\rm B},N_{\bar{\rm B}}}(\theta,\bar\theta)$.

We require the following four conditions for the properties 
of $k(\theta,\bar\theta)$:
\begin{align}
k(0,0) =& 0, 
\label{eq:ap:k0}
\\
\frac{ \partial^n }{ \partial \theta^n } k( 0,0 ) 
=& \xi_n N_{\rm B},
\label{eq:ap:k1}
\\
\frac{ \partial^n }{ \partial \bar\theta^n } k( 0,0 ) 
=& \bar\xi_n N_{\bar{\rm B}},
\label{eq:ap:k2}
\\
\frac{ \partial^{n+m} }{ \partial \theta^n \partial \bar\theta^m }
k( 0,0 ) =& 0,
\label{eq:ap:k3}
\end{align}
for positive integers $n$ and $m$.
Eq.~(\ref{eq:ap:k0}) is satisfied for probability distribution
functions normalized to unity.
Eqs.~(\ref{eq:ap:k1}) - (\ref{eq:ap:k3}) 
are Eqs.~(\ref{eq:k1}) - (\ref{eq:k3}) in the text.
All calculations in this Appendix are based only on these
constraints on $K(\theta,\bar\theta)$.

\subsection{Net proton number cumulants}

Using $K(\theta,\bar\theta)$, the net proton number 
cumulants are given by
\begin{align}
\langle (\delta N_p^{(\rm net)})^n \rangle_c
= \left( \frac{\partial}{\partial\theta} 
- \frac{\partial}{\partial\bar\theta} \right)^n
K(0,0).
\label{eq:ap:net}
\end{align}
To proceed the calculation of Eq.~(\ref{eq:ap:net}),
it is convenient to use the cumulant 
expansion of Eq.~(\ref{eq:ap:Kk})
\begin{align}
K( \theta,\bar\theta ) 
=& \sum_m \frac1{m!} \sum_F 
\left[ k(\theta,\bar\theta) \right]^m_c
\nonumber \\
=& 1 + \sum_F k(\theta,\bar\theta) 
+ \frac12 \sum_F ( \delta k(\theta,\bar\theta) )^2
\nonumber \\ &
+ \frac1{3!} \sum_F ( \delta k(\theta,\bar\theta) )^3
+ \frac1{4!} \sum_F ( \delta k(\theta,\bar\theta) )^4_c 
\nonumber \\ &
+ \cdots.
\label{eq:ap:cum}
\end{align}
Each term on the far right hand side defines 
each cumulant, $\sum_F[k(\theta,\bar\theta)]^m_c$, 
up to the fourth order,
with 
\begin{align}
\delta k(\theta,\bar\theta) &= k(\theta,\bar\theta) 
- \sum_F k(\theta,\bar\theta),
\\
\sum_F ( \delta k(\theta,\bar\theta) )^4_c 
&= \sum_F ( \delta k(\theta,\bar\theta) )^4 - 
3\left(\sum_F ( \delta k(\theta,\bar\theta) )^2\right)^2.
\end{align}

Because of Eq.~(\ref{eq:ap:k0}), all $k(\theta,\bar\theta)$ 
and $\delta k(\theta,\bar\theta)$ in a term in 
Eq.~(\ref{eq:ap:cum}) must receive at least one differentiation
so that the term gives nonzero contribution to 
Eq.~(\ref{eq:ap:net}).
This immediately means that the $m$-th order term in 
Eq.~(\ref{eq:ap:cum}) can affect Eq.~(\ref{eq:ap:net})
only if $m\le n$.

The first-order net-proton number cumulant, Eq.~(\ref{eq:Np1x}), 
is calculated to be
\begin{align}
\langle N_p^{\rm (net)} \rangle
=& ( \partial_\theta - \partial_{\bar\theta} ) K(0,0)
=\sum_F ( \partial_\theta - \partial_{\bar\theta} ) k(0,0)
\nonumber 
\\ 
=& \sum_F ( \xi_1 N_{\rm B} - \bar\xi_1 N_{\bar{\rm B}} )
= \langle \xi_1 N_{\rm B} - \bar\xi_1 N_{\bar{\rm B}} \rangle,
\label{eq:ap:Np1c}
\end{align}
with $\partial_\theta \equiv \partial / \partial\theta $
and $\partial_{\bar\theta} \equiv \partial / \partial\bar\theta $.
In the third equality in Eq.~(\ref{eq:ap:Np1c}), 
we have used Eqs.~(\ref{eq:ap:k1}) and (\ref{eq:ap:k2}).
The second-order relation, Eq.~(\ref{eq:Np2x}),
is obtained as follows:
\begin{align}
\lefteqn{ \langle ( \delta N_p^{\rm (net)} )^2 \rangle 
= ( \partial_\theta - \partial_{\bar\theta} )^2 K(0,0)}
\nonumber 
\\
&= \sum_F ( \partial_\theta - \partial_{\bar\theta} )^2 k(0,0)
+ \frac12 \sum_F ( \partial_\theta - \partial_{\bar\theta} )^2 
\left(\delta k(0,0)\right)^2
\nonumber 
\\ 
&= \sum_F ( \partial_\theta^2 + \partial_{\bar\theta}^2 ) k(0,0)
+ 2 \times \frac12 \sum_F \left[ ( \partial_\theta - \partial_{\bar\theta} )
\delta k(0,0) \right]^2
\nonumber 
\\ 
&= \xi_2 \langle N_{\rm B} \rangle 
+ \bar\xi_2 \langle N_{\bar{\rm B}} \rangle
+ \langle ( \xi_1 \delta N_{\rm B} 
- \bar\xi_1 \delta N_{\bar{\rm B}} )^2 \rangle.
\end{align}
To obtain the third line, we have used Eqs.~(\ref{eq:ap:k3})
and (\ref{eq:ap:k0}) for the first and second terms, 
respectively.
The factor two in the second term comes from the number of 
the outcomes of the application of the two derivatives to the two 
$\delta k(\theta,\bar\theta)$ in the second line.
Eqs.~(\ref{eq:ap:k1}) and (\ref{eq:ap:k2}) are used in 
the last equality.

Similar manipulations lead to Eqs.~(\ref{eq:Np3x})
and (\ref{eq:Np4x}).
We note that the relation,
\begin{align}
( \partial_\theta - \partial_{\bar\theta} )^4 \sum_F 
( \delta k(\theta,\bar\theta) )^4_c 
= 4! \sum_F \left[ ( \partial_\theta - \partial_{\bar\theta} )
k(\theta,\bar\theta) \right]^4_c,
\end{align}
makes the calculation for the fourth-order cumulant
more concise.

\subsection{Net baryon number cumulants}

To obtain Eqs.~(\ref{eq:NB1x}) - (\ref{eq:NB4x}),
we start from the following relation for the net baryon 
number cumulants,
\begin{align}
\langle (\delta N_{\rm B}^{\rm (net)})^n \rangle_c
& = \sum_F \left[ ( \xi_1^{-1} \partial_\theta 
- \bar\xi_1^{-1} \partial_{\bar\theta}) k \right]^n_c
\nonumber \\
& \equiv \sum_F 
\left[ \partial_\xi k \right]^n_c,
\label{eq:ap:B}
\end{align}
with $\partial_\xi = \xi_1^{-1} \partial_\theta 
- \bar\xi_1^{-1} \partial_{\bar\theta}$.
We suppress arguments in $K(0,0)$ and $k(0,0)$
throughout this subsection.

The manipulation of Eq.~(\ref{eq:ap:B}) for $n=1$ 
is trivial.
For $n=2$, Eq.~(\ref{eq:ap:B}) is calculated to be 
\begin{align}
\langle (\delta N_{\rm B}^{\rm (net)})^2 \rangle
&= \sum_F \left( \partial_\xi \delta k \right)^2 
= \frac12 \partial_\xi^2 \sum_F  \left( \delta k \right)^2 
\nonumber \\
&= \partial_\xi^2 K - \sum_F \partial_\xi^2 k
= \partial_{(1)}^2 K - \partial_{(2)} K
\nonumber \\
&= \langle ( \frac{\delta N_p }{\xi_1} 
- \frac{\delta N_{\bar p}}{\bar\xi_1}  )^2 \rangle
- \langle \frac{ \xi_2 }{ \xi_1^3 } N_p 
+ \frac{ \bar\xi_2 }{ \bar\xi_1^3 } N_{\bar p} \rangle.
\label{eq:ap:NB2}
\end{align}
In the second line, we introduced a symbol,
\begin{align}
\partial_{(n)} &= 
\frac{\xi_n}{\xi_1^{n+1}} \partial_\theta
+ (-1)^n \frac{\bar{\xi}_n}{\bar{\xi}_1^{n+1}} \partial_{\bar\theta},
\label{eq:ap:d1}
\end{align}
and used the relation,
\begin{align}
\partial_\xi^n k
&= \left( \frac1{\xi_1^n} \partial_\theta^n
+ \frac1{\bar\xi_1^n} \partial_{\bar\theta}^n \right) k
\nonumber \\
&= \left( \frac{ \xi_n }{ \xi_1^{n+1} } \partial_\theta
+ (-1)^n
\frac{ \bar\xi_n }{ \bar\xi_1^{n+1} } \partial_{\bar\theta} \right) k
= \partial_{(n)} k,
\end{align}
where we have used Eqs.~(\ref{eq:ap:k1}) - (\ref{eq:ap:k3}).
The last equality in Eq.~(\ref{eq:ap:NB2}) comes from 
the definition of $K$.

To proceed to $n\ge3$,
we first introduce the following notation,
\begin{align}
\partial_{(n,m)} &= 
\frac{\xi_n\xi_m}{\xi_1^{n+m+1}} \partial_\theta
+ (-1)^{n+m+1} (*\to\bar*),
\label{eq:ap:d2}
\\
\partial_{(n,m,l)} &= 
\frac{\xi_n\xi_m\xi_l}{\xi_1^{n+m+l+1}} \partial_\theta
+ (-1)^{n+m+l+2} (*\to\bar*),
\label{eq:ap:d3}
\end{align}
for positive integers $n$, $m$, and $l$.
$\partial_{(n_1,n_2,\cdots,n_i)}$ for $i>3$
is also defined as in 
Eqs.~(\ref{eq:ap:d1}), (\ref{eq:ap:d2}), and (\ref{eq:ap:d3}).
One easily finds 
i) $\partial_{(n,m,\cdots,l)}$ are invariant under 
the permutations of the subscripts, for example, 
$\partial_{(n,m,l)}=\partial_{(m,n,l)}$, and 
ii) when a subscript is one, it can be eliminated,
e.g., $\partial_{(n,m,1)}=\partial_{(n,m)}$, while 
$\partial_{(1)}=\partial_\xi$.
With this notation,
derivatives of $\delta k$ are written as
\begin{align}
\partial_\xi^n \delta k &= \partial_{(n)} \delta k,
\\
\partial_{(n)} \partial_{(m)} \delta k
&= \partial_{(n,m,2)} \delta k,
\\
\partial_{(n)} \partial_{(m)} \partial_{(l)} \delta k
&= \partial_{(n,m,l,3)} \delta k,
\end{align}
and so forth.

Using these relations, for example,
Eq.~(\ref{eq:ap:B}) for $n=3$ is calculated as
\begin{align}
\langle (\delta N_{\rm B}^{\rm (net)})^3 \rangle
&= \sum_F \left( \partial_\xi \delta k \right)^3
\nonumber \\
&= \partial_\xi^3 K 
- 3\sum_F ( \partial_\xi^2 \delta k )( \partial_\xi \delta k )
- \sum_F \partial_\xi^3 k 
\nonumber \\
&= \partial_{(1)}^3 K 
- 3\sum_F ( \partial_{(2)} \delta k )( \partial_{(1)} \delta k )
- \sum_F \partial_{(3)} k 
\nonumber \\
&= \partial_{(1)}^3 K 
- 3( \partial_{(2)} \partial_{(1)} K - \partial_{(2,2)} K )
- \partial_{(3)} K,
\end{align}
which leads to Eq.~(\ref{eq:NB3x}).
In the second and last equalities, we used 
\begin{align}
&\partial_\xi^3 K = \sum_F \left( \partial_\xi \delta k \right)^3
+ 3\sum_F ( \partial_\xi^2 \delta k )( \partial_\xi \delta k )
+ \sum_F \partial_\xi^3 k ,
\\
&\partial_{(n)}\partial_{(m)} K
= \partial_{(n,m,2)} K
+ \sum_F ( \partial_{(n)} \delta k )( \partial_{(m)} \delta k ).
\end{align}

A similar manipulation for $n=4$ leads to 
\begin{align}
\langle (\delta N_{\rm B}^{\rm (net)})^4 \rangle_c
=& \partial_{(1)}^4 K - 6 \partial_{(2)} \partial_{(1)}^2 K
+ 12 \partial_{(2,2)} \partial_{(1)} K 
\nonumber \\
&+ 3 \partial_{(2)}^2 K
- 4 \partial_{(3)}\partial_{(1)} K - 15 \partial_{(2,2,2)} K
\nonumber \\
& + 10 \partial_{(2,3)} K - \partial_{(4)} K,
\end{align}
which gives Eq.~(\ref{eq:NB4x}).

\section{Isospin density in final state}
\label{app:isospin}

In this Appendix, we demonstrate that the isospin density 
of nucleons in the final state of heavy ion collisions 
is suppressed owing to the abundant production of particles 
having nonzero isospin charges.

To simplify the calculation, we consider a gas composed 
of nucleons and pions in chemical equilibrium, and assume 
that pions and (anti-)nucleons obey Boltzmann statistics,
since this approximation does not alter the 
qualitative conclusion in this Appendix.
Under these assumptions, the ratios between the numbers of 
(anti-)protons and (anti-)neutrons
in a phase space are given in terms of
$\mu_{\rm I}$ and $T$ as 
\begin{align}
\frac{ N_p }{ N_n } = \frac{ N_{\bar n} }{ N_{\bar p} } 
= e^{\mu_{\rm I}/T} 
= \frac{ 1 - 2\alpha }{ 1+2\alpha },
\end{align}
with $\alpha = N_p / (N_p+N_n)$, and 
the ratio of the numbers of $\pi^+$ and $\pi^-$ is given by
\begin{align}
\frac{ N_{\pi^+} }{ N_{\pi^-} } = e^{2\mu_{\rm I}/T}.
\end{align}
With these relations, the total isospin in the phase
space is calculated to be
\begin{align}
N_{\rm I} 
=& \frac12 ( N_p - N_n - N_{\bar p} + N_{\bar n} )
+ N_{\pi^+} - N_{\pi^-}
\nonumber \\
=& \alpha \left( 
N_{\rm N} + N_{\bar{\rm N}} 
+ \frac4{1-4\alpha^2} N_{\pi_{\rm ch}} \right),
\label{eq:n_I}
\end{align}
with the number of charged pions 
$N_{\pi_{\rm ch}} =N_{\pi^+} + N_{\pi^-}$.

In the initial state of heavy ion collisions, the isospin 
asymmetry of the colliding heavy nuclei $\alpha_A$ is 
approximately $( N_n-N_p )/ (2 (N_p+N_n )) \simeq 0.1$.
Assuming that this isospin asymmetry equally distributes
along the rapidity direction in the final state,
one has $N_{\rm I}/ N_{\rm N}^{\rm (net)} \simeq \alpha_A$.
With Eq.~(\ref{eq:n_I}), one then obtains 
\begin{align}
\alpha \left( \frac{ N_N^{\rm (tot)} }{ N_N^{\rm (net)} }
+ \frac4{1-4\alpha^2} \frac{ N_{\pi_{\rm ch} }}{  N_N^{\rm (net)} }
\right)
\simeq \alpha_A.
\label{eq:ap:alpha}
\end{align}
The term in the parentheses
is larger than unity, and becomes larger as more charged pions 
and anti-nucleons are produced. 
Equation~(\ref{eq:ap:alpha}) thus shows that the value of 
$\alpha$ is more suppressed  than $\alpha_A$
owing to the production of these particles.
If the contribution of other particles
with nonzero isospin charges is taken into account,
the value of $\alpha$ is further suppressed.


\begin{thebibliography}{99}

\bibitem{RHIC}
  I.~Arsene {\it et al.}  [BRAHMS Collaboration],
  Nucl.\ Phys.\ A {\bf 757}, 1 (2005);
  B.~B.~Back {\it et al.} [PHOBOS Collaboration],
  Nucl.\ Phys.\ A {\bf 757}, 28 (2005);
  J.~Adams {\it et al.}  [STAR Collaboration],
  Nucl.\ Phys.\ A {\bf 757}, 102 (2005);
  K.~Adcox {\it et al.}  [PHENIX Collaboration],
  Nucl.\ Phys.\ A {\bf 757}, 184 (2005).

\bibitem{Asakawa:1989bq}
  M.~Asakawa and K.~Yazaki,
  Nucl.\ Phys.\  A {\bf 504}, 668 (1989).

\bibitem{Stephanov:2004wx}
  M.~A.~Stephanov,
  PoS {\bf LAT2006}, 024 (2006)
  [arXiv:hep-lat/0701002].

\bibitem{Aggarwal:2010wy}
  M.~M.~Aggarwal {\it et al.}  [STAR Collaboration],
  Phys.\ Rev.\ Lett.\  {\bf 105}, 022302 (2010)
  [arXiv:1004.4959 [nucl-ex]].

\bibitem{Mohanty:2011nm}
  B.~Mohanty  [STAR Collaboration],
  J.\ Phys.\ G {\bf 38}, 124023 (2011)
  [arXiv:1106.5902 [nucl-ex]];
  S.~Kabana  [for the STAR Collaboration],
  arXiv:1203.1814 [nucl-ex].

\bibitem{Bleicher:2011jk} 
  M.~Bleicher,
  arXiv:1107.3482 [nucl-th].

\bibitem{Koch:2008ia}
  V.~Koch,
  arXiv:0810.2520 [nucl-th].

\bibitem{Stephanov:1998dy}
  M.~A.~Stephanov, K.~Rajagopal, and E.~V.~Shuryak,
  Phys.\ Rev.\ Lett.\  {\bf 81}, 4816 (1998)
  [arXiv:hep-ph/9806219];
  Phys.\ Rev.\  D {\bf 60}, 114028 (1999)
  [arXiv:hep-ph/9903292].

\bibitem{Hatta:2003wn}
  Y.~Hatta and M.~A.~Stephanov,
  Phys.\ Rev.\ Lett.\  {\bf 91}, 102003 (2003)
  [Erratum-ibid.\  {\bf 91}, 129901 (2003)]
  [arXiv:hep-ph/0302002].

\bibitem{Stephanov:2008qz}
  M.~A.~Stephanov,
  Phys.\ Rev.\ Lett.\  {\bf 102}, 032301 (2009)
  [arXiv:0809.3450 [hep-ph]].

\bibitem{Athanasiou:2010kw}
  C.~Athanasiou, K.~Rajagopal and M.~Stephanov,
  Phys.\ Rev.\  D {\bf 82}, 074008 (2010)
  [arXiv:1006.4636 [hep-ph]].

\bibitem{Fraga:2011hi} 
  E.~S.~Fraga, L.~F.~Palhares and P.~Sorensen,
  Phys.\ Rev.\ C {\bf 84}, 011903 (2011)
  [arXiv:1104.3755 [hep-ph]].

\bibitem{Asakawa:2000wh}
  M.~Asakawa, U.~W.~Heinz, and B.~M\"uller,
  Phys.\ Rev.\ Lett.\  {\bf 85}, 2072 (2000)
  [arXiv:hep-ph/0003169].

\bibitem{Jeon:2000wg}
  S.~Jeon and V.~Koch,
  Phys.\ Rev.\ Lett.\  {\bf 85}, 2076 (2000)
  [arXiv:hep-ph/0003168].

\bibitem{Koch:2005vg} 
  V.~Koch, A.~Majumder and J.~Randrup,
  Phys.\ Rev.\ Lett.\  {\bf 95}, 182301 (2005)
  [nucl-th/0505052].

\bibitem{Ejiri:2005wq} 
  S.~Ejiri, F.~Karsch and K.~Redlich,
  Phys.\ Lett.\ B {\bf 633}, 275 (2006)
  [hep-ph/0509051].

\bibitem{Asakawa:2009aj}
  M.~Asakawa, S.~Ejiri, and M.~Kitazawa,
  Phys.\ Rev.\ Lett.\  {\bf 103}, 262301 (2009)
  [arXiv:0904.2089 [nucl-th]].

\bibitem{Friman:2011pf}
  B.~Friman, {\it et al.}, 
  Eur.\ Phys.\ J.\  C {\bf 71}, 1694 (2011)
  [arXiv:1103.3511 [hep-ph]].

\bibitem{Stephanov:2011pb} 
  M.~A.~Stephanov,
  Phys.\ Rev.\ Lett.\  {\bf 107}, 052301 (2011)
  [arXiv:1104.1627 [hep-ph]].

\bibitem{Gavai:2010zn}
  R.~V.~Gavai and S.~Gupta,
  Phys.\ Lett.\  B {\bf 696}, 459 (2011)
  [arXiv:1001.3796 [hep-lat]].

\bibitem{Schmidt:2010xm} 
  C.~Schmidt,
  Prog.\ Theor.\ Phys.\ Suppl.\  {\bf 186}, 563 (2010)
  [arXiv:1007.5164 [hep-lat]].

\bibitem{Mukherjee:2011td} 
  S.~Mukherjee,
  J.\ Phys.\ G G {\bf 38}, 124022 (2011)
  [arXiv:1107.0765 [nucl-th]].

\bibitem{Borsanyi:2011sw} 
  S.~Borsanyi, {\it et al.}, 
  JHEP {\bf 1201}, 138 (2012)
  [arXiv:1112.4416 [hep-lat]].

\bibitem{Bazavov:2012jq} 
  A.~Bazavov {\it et al.}  [HotQCD Collaboration],
  arXiv:1203.0784 [hep-lat].

\bibitem{Kitazawa:2011wh} 
  M.~Kitazawa and M.~Asakawa,
  Phys.\ Rev.\ C {\bf 85}, 021901R (2012)
  [arXiv:1107.2755 [nucl-th]].

\bibitem{Cleymans:1998fq}
  J.~Cleymans and K.~Redlich,
  Phys.\ Rev.\ Lett.\  {\bf 81}, 5284 (1998)
  [arXiv:nucl-th/9808030].

\bibitem{PDG}
  The Review of Particle Physics,
  K.~Nakamura, {\it et al.} (Particle Data Group), 
  J. Phys. G {\bf 37}, 075021 (2010).

\bibitem{Nonaka:2006yn}
  C.~Nonaka and S.~A.~Bass,
  Phys.\ Rev.\  C {\bf 75}, 014902 (2007)
  [arXiv:nucl-th/0607018].

\bibitem{BraunMunzinger:2003zd}
  P.~Braun-Munzinger, K.~Redlich and J.~Stachel,
  arXiv:nucl-th/0304013.

\bibitem{Schenke:2010nt} 
  B.~Schenke, S.~Jeon and C.~Gale,
  Phys.\ Rev.\ C {\bf 82}, 014903 (2010)
  [arXiv:1004.1408 [hep-ph]].
  
\bibitem{Pang:1992sk}
  Y.~Pang, T.~J.~Schlagel, and S.~H.~Kahana,
  Phys.\ Rev.\ Lett.\  {\bf 68}, 2743 (1992).

\bibitem{BraunMunzinger:2011dn} 
  P.~Braun-Munzinger, B.~Friman, F.~Karsch, K.~Redlich and V.~Skokov,
  Phys.\ Rev.\ C {\bf 84}, 064911 (2011)
  [arXiv:1107.4267 [hep-ph]];
  Nucl.\ Phys.\  A {\bf 880}, 48 (2012)
  [arXiv:1111.5063 [hep-ph]].

\bibitem{Karsch:2010ck}
  F.~Karsch and K.~Redlich,
  Phys.\ Lett.\  B {\bf 695}, 136 (2011)
  [arXiv:1007.2581 [hep-ph]].

\bibitem{Abelev:2008ab} 
  B.~I.~Abelev {\it et al.}  [STAR Collaboration],
  Phys.\ Rev.\ C {\bf 79}, 034909 (2009)
  [arXiv:0808.2041 [nucl-ex]].


\end{thebibliography}
\end{document}